%% file: main.tex
\documentclass[10pt,conference]{IEEEtran}
\usepackage{graphicx} 
\usepackage{comment}
\usepackage{enumerate}
\usepackage{xcolor}
\usepackage{subcaption}
\usepackage{paralist}
\usepackage{mdframed}
\usepackage{url}
\usepackage[hidelinks]{hyperref} 
\usepackage{booktabs} 
\usepackage{amsmath}
\usepackage{listings}

\newmdenv[
  backgroundcolor=gray!20,
  linewidth=0pt,
  innertopmargin=3pt,
  innerbottommargin=3pt,
  innerleftmargin=5pt,
  innerrightmargin=5pt,
]{graybox}

\usepackage{tikz}
\usetikzlibrary{shapes.geometric, arrows}

\usepackage{tabularx}
\usepackage{xspace}

\newcommand{\hpo}{hyperparameter optimization\xspace}
\newcommand{\approachhpo}{\texttt{InfPop}\xspace}
\newcommand{\llm}{LLM\xspace}
\newcommand{\hf}{HuggingFace\xspace}
\newcommand{\nvidia}{Nvidia\xspace}
\newcommand{\hfinference}{\texttt{HF$_{pl}$\xspace}}
\newcommand{\vLLM}{\texttt{vLLM\xspace}}
\newcommand{\hfsig}{HF\xspace}

\newcommand{\throughput}{throughput\xspace}

\newcommand{\ts}{t/s\xspace}

\title{
The Impact of Hyperparameters on Large Language Model Inference Performance: An Evaluation of vLLM and HuggingFace Pipelines% for Code-Completion Tasks
}
\author{\IEEEauthorblockN{Matias Martinez}
\IEEEauthorblockA{
Universitat Politècnica de Catalunya (UPC) - BarcelonaTech\\
Barcelona, Spain \\
matias.martinez@upc.edu}
}
\begin{document}

\maketitle

\newcommand{\rqshapespace}{RQ1: What is the shape of the throughput landscape across the hyperparameter space defined by the batch size and the number of GPUs on \vLLM{} inference?}

\newcommand{\rqonlineinferenceHFGPU}{RQ2: For online inference, how does throughput vary with different numbers of GPUs (hardware scaling)?}

\newcommand{\rqBatchInference}{RQ3: For batched inference, how does throughput vary with the batch size?}

\newcommand{\rqOlderHardare}{%RQ4: What is the impact of using older -previous-generation- GPU hardware?
%RQ4: What is the impact of the GPU hardware ?
RQ4: How does \throughput{} vary between different GPU models during inference?
%RQ4: How does the \throughput{} of the inference tasks vary across different GPU models?
%How does \throughput{} of the inference vary according to the GPU model used?
}

\newcommand{\rqHPO}{RQ5: To what extent does applying Hyperparameter Optimization improve \throughput{} in \llm{}?}

\begin{abstract}
   % The current wave of open source large language models (LLM) allow developers to create AI-based solutions while controlling different aspects such as privacy, and compliance, providing them governance and ownership of the deployment process. 
    
   % To interact with these LLMs, inference engines are required, which are in change of loading the model (i.e., its weights) on the available resources and to then execute the inference (i.e., given the query, execute the model to produce an output).

   % The speed of inference or performance of the LLM inference critical for real-time applications,as it has to compute millions or billions of floating point operations per each inference. 

  %  Recently, inference engines, such  which implement novel inference mechanism, as \vLLM{}  have emerged to provide state-of-the-art performance.

   % In this paper, we study the performance (in particular, the \throughput{}) of 20 LLMs using two popular inference libraries: \vLLM{} and \emph{pipelines} from HuggingFace.
   % In particular,  we focus on how hyperparameters, that developers have to set, affect on the  performance of inference.

  %  Our results show that the \throughput{} landscapes are irregular (e.g., have peaks), involving that hyperparameter optimization is capital to achieve maximum performance.
  %  We also present a study showing that applying hyperparameter optimization in a scenario of upgrading or downgrading the GPU model used for inference helps to increase \throughput{}, on average up to 13.7\%. 

The recent surge of open-source large language models (LLMs) enables developers to create AI-based solutions while maintaining control over aspects such as privacy and compliance, thereby providing governance and ownership of the model deployment process.
To utilize these LLMs, inference engines are needed. These engines load the model’s weights onto available resources, such as GPUs, and process queries to generate responses.
The speed of inference, or performance, of the LLM, is critical for real-time applications, as it computes millions or billions of floating point operations per inference.
Recently, advanced inference engines such as \vLLM{} have emerged, incorporating novel mechanisms such as efficient memory management to achieve state-of-the-art performance.
In this paper, we analyze the performance, particularly the \throughput{} (tokens generated per unit of time), of 20 LLMs using two inference libraries: \vLLM{} and HuggingFace's \emph{pipelines}. We investigate how various hyperparameters, which developers must configure, influence inference performance.
Our results reveal that \throughput{} landscapes are irregular, with distinct peaks, highlighting the importance of hyperparameter optimization to achieve maximum performance.
We also show that applying hyperparameter optimization when upgrading or downgrading the GPU model used for inference can improve \throughput{} from HuggingFace pipelines by an average of 9.16\%  and  13.7\%, respectively.
    
\end{abstract}

\begin{comment}
https://conf.researchr.org/track/icse-2025/icse-2025-research-track
    
    i) Novelty: The novelty and innovativeness of contributed solutions, problem formulations, methodologies, theories, and/or evaluations, i.e., the extent to which the paper is sufficiently original with respect to the state-of-the-art.

ii) Rigor: The soundness, clarity, and depth of a technical or theoretical contribution, and the level of thoroughness and completeness of an evaluation.

iii) Relevance: The significance and/or potential impact of the research on the field of software engineering.

iv) Verifiability and Transparency: The extent to which the paper includes sufficient information to understand how an innovation works; to understand how data was obtained, analyzed, and interpreted; and how the paper supports independent verification or replication of the paper’s claimed contributions. Any artifacts attached to or linked from the paper will be checked by one reviewer.
\end{comment}

\section{Introduction}
%\todo{cite in the intro \cite{bai2024beyond}\cite{li2024llm}}

%Yarally2023BatchingGreenAI
%The batch size (i.e., the number of input data
%samples that are processed at one time during inference) is
%one of the most important hyperparameters to tune during the
%training phase. It has implications on the model accuracy and
%generalisability [5], training times and parallelisability [13],
%etc.

%Xhou et al. \cite{zhou2024survey} survey on efficient inference  for LLMs, which  targets different aspects, from providing primary causes of
%the inefficient LLM inference to  defining comprehensive taxonomies of, for instance, inference engines.
%In particular, they analyze features from inference engines including the two we consider, \hfinference and \vLLM{}
%They also conducted experiments such as the evaluation of the speed-ups achieved
%by employing uantization techniques on two models LLaMA-2-7B and 13B using two inference frameworks, TensorRT-LLM\cite{}  and LMDeploy\cite{}. %Recent advances on inference are also discussed in \cite{li2024llm}
%They used a single single NVIDIA A100 GPU and 
%\cite{miao2023towards} survey

%\cite{shi2024efficient} to cite, importan vision paper.

%\todo{\cite{yang2024robustness} no mention any hpo work on LLM}

%\todo{\cite{miao2023towards} survey inference, different to the other}

%\todo{check survey LLM for codegen\cite{jiang2024survey}}

%\todo{\cite{ghadesi2024causes} hyperparameter in SO including HF post.}

Large language models (LLM) have revolutionized the manner in which developers build \emph{intelligent} software solutions, affecting several areas, including software engineering (\cite{zhang2023survey,zhang2023unifying,jiang2024survey,ghaemi2024transformers}).
One of the primary drivers of this revolution is the emergence of \emph{open-source} LLMs, which offer publicly accessible architecture, code, checkpoints (model weights) and, eventually, training data.
Open-source LLMs enables developers to create AI-based solutions while maintaining control over aspects such as privacy and compliance, and, at the same type, providing governance and ownership of the model deployment process.
Platforms such as \hf{} (\hfsig) have contributed to the success of open-source AI by providing, for instance, more than 700.000 models for download, including LLMs such as CodeLlama~\cite{roziere2023codellama}.

Inference engines (or frameworks) are software components that make LLMs operable.
They have as main responsibility to load the model's weighs into the available devices (GPU and/or CPU), and to make the \emph{inference} i.e., to generate a prediction/output/response based on input data.
\hf{}, for instance, provides inference mechanisms in its \texttt{Transformer} library~\cite{wolf-etal-2020-transformers}, such as the \texttt{pipelines}, which are objects that abstract complex code used for inference (by wrapping other libraries such as PyTorch) and offer a simple API dedicated to several tasks, including text generation.

%The appearance of open-source \llm{} encourages this latter strategy, as a developer can download a model, modify it (e.g. applying fine-tuning, pruning, quantization, or other techniques) and then deploy it in a desired infrastructure as those mentioned.
%For example, \hf{} (\hfsig), the platform that drives open-source initiatives on AI,  has more than 700.000 models available for download.
%To applying this strategy, the developers need an inference engine, which is a piece of software that has as main responsibility to load the model into the available devices (GPU and/or CPU), and then to query it (i.e., to perform the inferences).
%Among all the functionalities and features provided by \hf, there are two that simplify the integration of \llm{} into apps, available on library \texttt{Transformer}~\cite{wolf-etal-2020-transformers}.
%The first one, \texttt{AutoModel.from\_pretrained(<model\_id>)}, downloads and loads a model on the available devices.
%Then, \hf{} provides different methods for performing inference on that loaded model, such as \texttt{pipelines}, which offer a simple API dedicated to several tasks, including text generation.
%For instance, \hf{} provides \texttt{pipelines}, which are objects that abstract complex code used for inference (which wraps other libraries such as PyTorch) and offer a simple API dedicated to several tasks, including Text-Generation.
%\hf{} pipelines simplifies the manner to integrate and use \llm{} within software application.  

A key factor for the success of an inference engine is \emph{performance}, i.e., the speed of serving, because it directly impacts the user experience of AI-powered systems and on the cost (since LLMs require specialized hardware such as GPUs).
A measure of LLM performance is \emph{\throughput{}}, expressed in tokens generated by an \llm{} per unit of time.
Increasing the throughput of LLM inference, and therefore reducing the cost per request, is becoming primordial~\cite{kwon2023efficient}.

Recently, state-of-the-art inference engines have emerged with the goal of improving inference performance. 
One of them is \vLLM{}~\cite{kwon2023efficient}, launched in June 2023, which implements a memory-efficient inference engine for \llm{}.
Inference engines such as vLLM or these from HuggingFace can be integrated into applications with a few lines of code, as \autoref{lsthf1} shows.

However, the inference engine has hyperparameters that developers need to set by the developers.
For example, Listing \ref{lsthf1} shows a real utilization of \hf{} pipelines\footnote{\url{https://github.com/adit-copilot/EnchantedQuest/blob/48d6256c482abc40a8c916c6c8d1a7c5edb2c6fa/retrival/qa_pipeline.py\#L46}}, which first loads the model (it has one hyperparameter \texttt{device\_map}) and then creates a pipeline with the model as a parameter and some hyperparameters (e.g, \texttt{max\_new\_tokens}). 
Listing~\ref{lstvll1} shows  real code (but simplified)  that uses \vLLM{}\footnote{\url{https://github.com/tencent-ailab/persona-hub/blob/5e9b7d0a29eff2e2eaf01cc8c74d4abb731fb4a0/code/vllm\_synthesize.py\#L37}}. 
The first line loads the model (with hyperparameter \texttt{tensor\_parallel\_size}), the second performs inferences from a list of prompts (receives two hyperparameter \texttt{temperature} and  \texttt{top\_p}).

% //Snippet 1: App https://github.com/tencent-ailab/persona-hub
\begin{lstlisting}[language=Python, caption=\mbox{\hf{} pipeline},
 basicstyle=\tiny, %or \small or \footnotesize etc.
label=lsthf1,
float=tp,
  floatplacement=tbp,
]   
    model = AutoModelForCausalLM.from_pretrained(model_name, device_map='auto',[...])
    return pipeline("text-generation", 
        model=model, 
        max_new_tokens=100,
        model_kwargs={"torch_dtype": torch.bfloat16} [...] )
\end{lstlisting}

\begin{lstlisting}[language=Python, caption=\mbox{\vLLM{}},
 basicstyle= \tiny, %or \small or \footnotesize etc.
label=lstvll1,
float=tp,
  floatplacement=tbp,
]
    llm = LLM(model=model_path, tensor_parallel_size=4)
    #tensor_parallel_size based on the GPUs you are using
    outputs = llm.generate(prompts, SamplingParams(temperature=0.6, top_p=0.95, [...]))
\end{lstlisting}

%From the listings we observe that both libraries have \emph{hypermarameters}.
%Listing \ref{lstvll1} shows three, one used on model load  (\texttt{tensor\_parallel\_size=4}) and other two are used during inference (\texttt{temperature=0.6} and  \texttt{top\_p=0.95}).

%, inparticular, those that have a direct incidence in \throughput{}. Other }} 
There are hyperparameters that have a direct incidence in \throughput{} while others such as \texttt{temperature} direct affects the output from an \llm{}.
In this paper, we focus on the former.
One of them is 
\texttt{tensor\_parallel\_size} from \vLLM{}, which indicates the number of GPUs to use for distributed execution with tensor parallelism. %\footnote{\url{https://github.com/vllm-project/vllm/blob/main/vllm/entrypoints/llm.py\#L44C9-L44C29}}
The value of this hyperparameter needs to be set by the developer according to the resources available in the deployment infrastructure.
In the code from \autoref{lstvll1}, \texttt{tensor\_parallel\_size} is hard-coded to 4 and includes the comment:
\emph{``\#tensor\_parallel\_size based on the GPUs you are using''}.
%Both \texttt{from\_pretrained} and \texttt{pipeline} form 
%\hf{} inference mechanisms (e.g., \emph{pipelines}) have a hyperparameter, named \texttt{device\_map},  which indicates the distribution of the LLM in available devices (GPU, CPU).
%In Listing\ref{lsthf1}, \texttt{device\_map} receives the value \texttt{auto}, which means that \hf{}  automatically determines the deployment of the model depending on the available -and visible- resources. 
%\footnote{CUDA\_VISIBLE\_DEVICES}) 

Another hyperparameter that affects \throughput{} is \emph{batch size}, which refers to the number of input instances processed simultaneously by the model in a single inference call. 
According to \cite{Yarally2023BatchingGreenAI}, the batch size is one of the most important hyperparameters to tune during the training phase, affecting, for example, accuracy and training time.
In inference, increasing batch size typically increases throughput~\cite{kwon2023efficient}.
However, determining the batch size is challenging as is subject to limitations of the memory capacity of the hardware (e.g., it is not able to allocate the memory for numerous inputs on a large batch).
On the other hand, a low value can lead to hardware being underutilized.

In this paper, we study how hyperparameters of inference engines affect the \throughput{} of \llm{}.
In particular, we examine the throughput landscape defined by the spaces of two hyperparameters, the number of GPUs to be used, and the batch size.
We study LLM inference in the context of the code completion task, which is implemented by products such as Microsoft GitHub Copilot \cite{Ziegler2022Copilot,Ziegler2024CopilotACM}. %implement this task.
As noted by \cite{white2023navigating}, copilot efficiency is essential because LLM inference is costly and unsustainable on a massive scale.
In our experiment, we study 20 open-source \llm{}, all from key players in the AI industry such as Meta, Google, Microsoft, Mistral, and HuggingFace, and two popular libraries that enable inference: HuggingFace Transformers \cite{wolf-etal-2020-transformers} and \vLLM{} \cite{kwon2023efficient}.

Our results reveal that the \throughput{} landscapes are irregular, with distinct peaks, underscoring the importance of hyperparameter optimization for maximizing inference performance.
Motivated by these findings, we conduct an experiment on hyperparameter optimization.
We examined two use cases focusing on changes in the inference infrastructure, such as upgrading or downgrading the GPU model.
Our experiment shows that adjusting the number of GPU hyperparameters and/or batch size using our tool \approachhpo{} (based on Hyperopt~\cite{pmlr-v28-bergstra13Hyperopt}) 
in Hugging Face inference improves throughput by an average of 9.16\% when upgrading the GPU model (from \nvidia{}  V100 to A100) and by 13.7\% when downgrading (from \nvidia{}  A100 to V100).
%in HuggingFace inference improves \throughput{} on average 9.16\% when there is an upgrade of the GPU model (from  \nvidia{} V100 to A100) and an improvement of 13.7\% when there is a downgrade (from \nvidia{} A100 to V100).

All data, plots, and scripts are available in our Appendix~\cite{appendix}.
The paper continues as follows.
Section \ref{section:rq} presents the research question we investigate.
Section \ref{sec:methodology:experimentalprotocol} describes the research protocol.
Section \ref{sec:results} presents the results.
Section \ref{sec:discussion} presents the discussion.
Section \ref{sec:relatedwork} discusses related work. 
Section \ref{sec:conclusion} presents the conclusions and future work.

\section{Research questions}
\label{section:rq}

The research questions that guide this experiment are the following:
%\medskip
\begin{graybox}
\rqshapespace
\end{graybox}

For each \llm{} under study, we visually inspect the throughput distribution within the hyperparameter space defined by the number of GPUs and the batch size to analyze its shape.
In particular, we search for:
\begin{inparaenum}[\it 1)]
\item[] peaks, 
\item[] flat regions, or
\item[] valleys.
\end{inparaenum}

This analysis gives us the first sing to know whether hyperparameter optimization should be applied to maximize or minimize a metric, such as \throughput{}.
For example, a peak in the landscape is a sign that there is a set of hyperparameters that maximize throughput.

%Once we perform the visual analysis, in the subsequent research questions we apply a series of numerical analysis in order to quantify the impact of hyperparameter changes on the \throughput{} under certain conditions, including type of inference, inference engine, and hardware used.

%\medskip
\begin{graybox}
\rqonlineinferenceHFGPU
\end{graybox}

The purpose of this question is to study the impact of the number of GPUs on \throughput{}.
This question could help us to know, for example, whether having more GPUs in the inference infrastructure (called \emph{GPU scaling}) allows an inference engine to improve \throughput{}.
%The number of GPUs is one hyperparameter of some inference engine, including the two, \vLLM{} and \hfinference{}, presented in Section \ref{sec:backgroud}.
%Here, we study the relation between that hyperparameter space and the \throughput{}.

In this research question, we focus on \emph{online inference}, 
each query done to the model (i.e., a request) through the inference library includes just one input. As a result, the model generates one output (or eventually more than one if applying, for example, beam search) for that single input.

%The alternative to \emph{online inference}  is \emph{batching}, which  consists in sending multiple inputs in a single query. 
%The inference produces, as a response to that query, one or more outputs of each of the inputs included in the query.
%We study the impact of batching in RQ3.

%Figure \ref{fig:inferenceOnlineBatch} shows the differences between online and batching inference processes.
%\input{figInference}.
%\todo{explain}.

%\medskip
\begin{graybox}
\rqBatchInference
\end{graybox}

This research question focuses on the \emph{batch} inference. 
In contrast to the previous one, where the query to the \llm{} consists of one \emph{single} input, the batching considers $n$ inputs in a single query.
This $n$ is the \emph{batch size}, which is an hyperparameter of the inference processes. 
Choosing an appropriate batch size $n$ is challenging.
On the one hand, a high batch size value would lead to throughput improvements but also may lead to out-of-memory.
On the other hand, a low value may lead to resources (e.g., GPUs) being underutilized.
Here, we study the relation between the space of the hyperparameter batch size and the \throughput{} from inference engines.

%Choosing an appropriate batch size $n$ is challenging.
%On the one hand, a high value of batch size would lead optimize the resource (GPU) utilization and, consequently, increase the throughput, expressed in tokens per second.
%However, at the same time, a high batch size may lead to an out-of-memory exception: for example, the GPU is not able to allocate the memory of numerous inputs.
%On the other hand, a low value may lead to resources (e.g., GPUs) being underutilized.

%\medskip
\begin{graybox}
\rqOlderHardare
\end{graybox}
%\medskip

Each new generation of GPUs boasts improved performance including \throughput{} \cite{Choquette2021NvidiaA100}.
In this research question, we measure the variation in  \throughput{} across two GPU models, one representing a recent and widely used model, and another model from a previous generation.

%\medskip
\begin{graybox}
\rqHPO
\end{graybox}
%\medskip

The goal is to present an use case in which \hpo{} of \llm{} inference produces improvements in \throughput{}.
In particular, we apply \hpo{} (using our approach \approachhpo{}) in cases where there are changes on the hardware infrastructure used for inference.
Thus, \hpo{} finds a new  hyperparameter values that maximize \throughput{} on the new hardware.

\section{Experimental protocol}
\label{sec:methodology:experimentalprotocol}

In order to answer the research questions, we conduct an experiment that consists of executing the code-completion task for a set of incomplete programs. 
That completion is done by doing inference calls to a \llm{} using an inference engine. 
The experiment is designed as follows.
%The inference of the missing code is done via an inference engine, which loads the model on the available devices (e.g., CPU, GPUS) and executes the queries on it.

%This section introduces all steps to carry on our experiment, and continues as follows.
%Section \ref{}
%Section \ref{} presents the criteria to select the \llm{} to evaluate.
%Section 
%\todo{to complete}
%We detail them in Section %\ref{sec:methodology:dataset}

\subsection{Selection of the inference engines to evaluate}
\label{sec:methodology:selectionInferenceengine}

%% https://github.com/search?q=%22from+vllm+import+LLM%22+language%3APython+&type=code

We evaluate inference using two inference engines. %\hf{} inference and \vLLM{}.

\subsubsection{Inference  from \hf{} Transformers library}
We select it for the following reasons.
First, it offers  simple, easy to use, and concise APIs to perform inference on \llm{}.
For example, \emph{pipelines} abstracts most of the complex code from the library, and provides a simple API dedicated to several tasks, including Text Generation. An example for this pipeline  is shown in \autoref{lsthf1}.
% shows a pipeline for the ``text generation'' tasks, which englobe our specific task (code completion).
Internally, that code uses a pipeline \texttt{TextGenerationPipeline}. 
%, included in the HF Transformers library\footnote{\url{https://huggingface.co/docs/transformers/en/main\_classes/pipelines\#transformers.TextGenerationPipeline}}. 
In the remainder of the paper, we will refer to the HuggingFace pipeline for text generation it as \hfinference{}.
Secondly, \hf{} provides thousads of models which can be used with  \hfinference{} for inference.
Lastly, HF Transformers has is a relevant library in the  the scene of open-source on AI: it has $\approx$130.000 stars on GitHub.
%We choose to study the inference method provided \hf{} for the relevance that it has in the scene of open-source on AI: HF Transformers has $\approx$130.000 stars on GitHub.

\subsubsection{\vLLM{}}
 originaly materialized to implement  \emph{PagedAttention}~\cite{kwon2023efficient} a technique  for efficient management of attention key and value memory of transformer architecture. 
%Other key features from \vLLM{} include:
%\begin{inparaenum}[\it 1)]
 %   \item Continuous batching of incoming requests,
 %   \item A wide range of Quantization techniques (GPTQ, AWQ, SqueezeLLM, FP8 KV Cache), and
 %   \item optimized CUDA kernels.
%\end{inparaenum}
We choose to study \vLLM{} for different reasons, including:
\begin{inparaenum}[\it a)]
   \item It provides state-of-the-art inference performance on \llm{}~\cite{kwon2023efficient}.
	\item introduces a novel memory management (PagedAttention~\cite{kwon2023efficient}), then integrated to other inference engines \cite{li2024llm} such as TensorRT-LLM \cite{tensorRTLLM}.
    \item It integrates with \hf{} models, enabling it to work with state-of-the-art LLMs.
    \item Popularity: more than 23.000 Github stars since its launch (June 2023).
    \item Integrated with other \llm{} libraries and frameworks such as LangChain\footnote{\url{https://python.langchain.com/v0.2/docs/integrations/llms/vllm/}}. %and  SkyRocket\footnote{\url{https://docs.vllm.ai/en/latest/serving/run_on_sky.html}}. % which facilitates its adoption.
\end{inparaenum}

\subsection{Selection criteria of the \llm{} to evaluate}
\label{sec:methodology:selectionLLMs}

\input{tables/tabLLMInfo}
%During recent years, researchers and companies have released numerous deep learning models capable of understanding and generating source code~(e.g., \cite{jiang2024survey,yang2024ecosystemLLMonCode,zhang2023survey,zhang2023unifying}).
%In this work, we focus on autoregressive models, particularly those based on the Transformer architecture~\cite{vaswani2017attention}. 
%This architecture is highly effective for code auto-completion tasks due to their ability to generate sequences token-by-token, maintaining context and dependencies \cite{Svyatkovskiy2020IntelliCode}.

%The model selection criteria are the following:
We select a model if all the following criteria are met.
\begin{itemize}
    \item Model based on the Transformer architecture~\cite{vaswani2017attention}, which is effective for code completion tasks~\cite{Svyatkovskiy2020IntelliCode}. %due to their ability to generate sequences token-by-token, maintaining context and dependencies \cite{Svyatkovskiy2020IntelliCode}.
    \item As our study focuses on large language models (LLM), the model must have $\approx >$1 billion parameters.%\footnote{Determining when a model is ``large" is quite arbitrary, as there is no strict size threshold on the number of parameters, architectures or amount of data used in training.}
    %\item fully or partially trained with source code.
    \item Evaluated on HumanEval~\cite{chen2021evaluating} to demonstrate their competence in the code generation task.
    \item Model's weights (checkpoints) are available on the HuggingFace hub\footnote{HuggingFace model hub: \url{https://huggingface.co/models}}.
    \item Model's architecture implemented in the \texttt{HuggingFace Transformers} library\footnote{Model architectures implemented in HuggingFace transformer library: \url{https://github.com/huggingface/transformers/tree/14ff5dd962c1bd0a4e3adaac347ba396d8df5add/src/transformers/models} and \url{https://huggingface.co/docs/transformers/}}~\cite{wolf-etal-2020-transformers}. % or this library is able to load models that it does not include.\footnote{Possible using option \texttt{trust\_remote\_code=True} in method \texttt{from\_pretrained} from class \texttt{AutoModelForCausalLM}.}
    %\item In the \hf{} site of the model, it  is tagged with ``Text-Generation"\footnote{\url{https://huggingface.co/models?pipeline_tag=text-generation}}  tag. This ensures that it can be used with the HF pipeline.
    \item Model tagged with ``Text-Generation" tag\footnote{\url{https://huggingface.co/models?pipeline_tag=text-generation}} in the \hf{} site for ensuring usage with HF pipeline.
    \item Model supported by the vLLM library.
\end{itemize}

To select the models to study, we first inspect the vLLM documentation and its Github repository to detect the supported models.
As \vLLM{} uses HuggingFace Transformers library, this strategy allows us to select models supported by both the \hfinference{}  and \vLLM.
For each supported model, we read its documentation, related paper, HuggingFace model card, and the dashboard Paper-on-Code\footnote{\url{https://paperswithcode.com/sota/code-generation-on-humaneval}} to check all mentioned criteria.

Table \ref{tab:model_data} shows the 20~\llm{} that meet the criteria mentioned above and work in our infrastructure.
It shows the numbers of parameters, which range from 1.3 to 70 billions, and the HumanEval (pass@1) reported by the models's authors. 
We include \llm{} from key players in the AI industry, such as Meta, Google, Microsoft, Mistral, and HuggingFace\footnote{HuggingFace participates via BigCode:  \url{https://www.bigcode-project.org/}} (\cite{li2023starcoder,lozhkov2024starcoder}).
Other models are discarded because they do not meet some of the criteria. For example, 
Moisaic~\cite{MosaicML2023Introducing} has not been evaluated in HumanEval;
CodeGen \cite{nijkamp2022codegen} and Incoder~\cite{fried2022incoder} are not supported by vLLM (last check June 2024).
Microsoft-Phi-3 was later discarded because it failed in our infrastructure. %we failed to deploy it due to an installation problem with its dependency FlashAttention\footnote{https://github.com/Dao-AILab/flash-attention}.

\subsection{Evaluation benchmark}
\label{sec:methodology:dataset}

HumanEval \cite{chen2021evaluating} is a dataset of 164 handwritten programming problems proposed by OpenAI to evaluate functional correctness and measure the problem-solving capabilities of their code-based models.
%In particular, \cite{chen2021evaluating} evaluated the ability of their model Codex to generate correct Python functions given as input a \texttt{docstring} i.e., a description of the program to generate.
%Programming tasks in the HumanEval dataset assess language comprehension, reasoning, algorithms, and simple mathematics.
%We consider HumanEval as it has been one of the most used benchmarks for code generation (e.g. code \llm{} from Meta, Google and Microsoft use it).
We consider HumanEval as it has been used by  big AI players (e.g., OpenAI, Meta, Google, Microsoft) to evaluate the capaibilities of their LLMs.

Researchers created new benchmarks  from HumanEval, to evaluate other tasks beyond function generation.
For example, Fried et al.~\cite{fried2022incoder} and Bavarian et al.~\cite{bavarian2022efficient} evaluated code \emph{infiling} task, which consists of predicting missing text spans that are consistent with the preceding and subsequent code.

To evaluate the code completion task, we select  \emph{Single-line infilling}\footnote{\url{https://github.com/openai/human-eval-infilling/blob/88062ff9859c875d04db115b698ed4b0f0395170/data/HumanEval-SingleLineInfilling.jsonl.gz}}~\cite{fried2022incoder}.
They created it by marking out each non-blank line of code in the canonical function implementation from each HumanEval problem. %, creating N examples for a function with N non-blank lines per problem.
In total, \emph{Single-line infilling} contains 1033 examples. 
%Each of these examples has:
%\begin{inparaenum}[\it a)]
  %  \item a masked line,
   % \item the code that precedes the masked line,
    %\item the code that is subsequent to the masked line.
%\end{inparaenum}
We use this bechmark in a slightly different manner:
we query a model to generate code just based on the preceding code, and we ignore the subsequent code as code completion is not infilling.
%We do not asset the correctness of the generated code because it is not within the scope of this paper.

\subsection{Hardware used}
\label{sec:methodology:hardware}

We use two high-performance computation nodes available in our host infrastructure: two CPU Intel Xeon E5-2698 and 512 GB of RAM.
The first one is equipped with eight \nvidia{} A100 GPUs with 40 GB ~\cite{Choquette2021NvidiaA100} .
We choose this node because that GPU has been extensively used to train and evaluate state-of-the-art LLMs considered in our study. 
The second node is equipped with a GPU model from a previous generation: \nvidia{} V100\footnote{\nvidia{} V100 whitepaper: \url{https://images.nvidia.com/content/volta-architecture/pdf/volta-architecture-whitepaper.pdf}}, which is the first Tensor Core GPU introduced by \nvidia{} designed specifically for deep learning.
The node is equipped with eight V100 with 32 GB.
We choose to consider V100 as a ``previous generation'' GPU because it has been used to train and evaluate some considered LLMs (e.g. Bloom~\cite{workshop2022bloom}).

%Tesla V100-SXM2-32GB 
%https://www.nvidia.com/en-us/data-center/v100/

%Tesla V100-SXM2-32GB 
%https://www.nvidia.com/es-es/data-center/tesla-v100/

\subsection{Executing Inference trials on the Single-line infilling}
\label{sec:methodology:vars}
\label{sec:methodology:inference}

We call a \emph{trial} to the execution of the code completion tasks for all items from the  \emph{Single-line infilling} benchmark by querying one \llm{}  $m$, using an inference engine $e$ ($e \in [\hfinference{},\vLLM])$, executed on $n$ GPUs  ($n \in [1, 2, 4, 8]$)  of model $g$ ($g \in [A100, V100]$), with a batch size $b$ ($b \in [1, 2, 4, 8, 16, 32, 64, 128, 256, 1028]$). 
We run \emph{trials} resulting from all the combinations between $m$, $e$, $n$, $g$ and $b$.
Note that, for each \emph{trial}  with batch size $b$, we create $x$ inference invocations, each with $b$ elements from the benchmark ($x$ results from the division between the size of the benchmark and $b$). 
%For example, if the batch size is 4, we create 258 batches (1033 / 4) and the inference is then invoked for each of them.
In \vLLM{}, we  modify the parameter of  the  constructor of class \texttt{LLM}. 
\texttt{tensor\_parallel\_size} for pasing the number of GPUs to be used, and pass $b$ queries to the method  \texttt{generate} from that class, where $b$ is the batch size of the trial.
In \hfinference{},  we keep \texttt{device\_map="auto"} from the pipeline and, at the same time, we restringe the visible GPUs by changing the  evironmental variable \texttt{CUDA\_VISIBLE\_DEVICES}.
For example, for a trial that uses 2 GPUs, we set the value \texttt{0,1}. Then, the other GPUs ($2\dots8$) are not visible.
This strategy allows the pipeline, thanks to the \texttt{auto} configuration provides by the HF Accelerate library, to determine automatically the model deployment  on the available (and visible) devices.
For the batch size, we pass the value $b$ via hyperparameter \texttt{batch\_size} from the pipeline.

%https://huggingface.co/docs/accelerate/v0.32.0/en/concept_guides/big_model_inference
%(https://huggingface.co/transformers/main_classes/pipelines.html#pipeline-batching

We keep the default values for the other hyperparameters from the inference engine,  except for two.
First, we indicate the inference engine to generate up to 5 tokens.
Second, we unify the floating-point precision used.
All models from \autoref{tab:model_data} use by default half-precision floating point (FP16/float16) with the exception of one (bigcode/starcoder2-3b) that uses full-precision floating point (FP32/float32).
To ensure a fair comparison between models and engines, we explicitly set the hyperparameter precision to FP16 in both inference engines as is the value from the large majority.
%\todo{versions of vllm and transformer?}
For each trial, we calculate \throughput{} as the number of tokens generated during the trial (composed by $x$ inference invocations) divided by the execution time of the trial in seconds (\ts{}).

\subsection{Data analysis to answer the research questions}
\label{sec:methodology:rq}

Once executed all trials, we apply the following analysis to answer each research question.

%\subsubsection{\rqshapespace}
\subsubsection{RQ1}
We plot the \throughput{} landscapes using the Python library matplotlib. There is a plot for each combination of:
\begin{inparaenum}[\it a)]
    \item hardware,
    \item inference engine, and 
    \item \llm{}.
\end{inparaenum}
We then inspect the shape of all the landscapes to select the representative cases for discussion.

%\subsubsection{\rqonlineinferenceHFGPU}
\subsubsection{RQ2}

We select the trials with batch size equal to one.
Then, for each \llm{}, we plot the \throughput{} evolution with different numbers of GPUs, and report the number of GPUs that maximize \throughput{}. 
We test the hypothesis \emph{$H_0$:~there is no statistically significant difference in throughput across different numbers of GPUs} by appling an analysis of variance (ANOVA).
In this test, each group  corresponds to the trial with the same number of GPUs used on inference (1, 2, 4 and 8).
Finally, we compare \throughput{} between the two inference engines, \vLLM{} and \hfinference{}.

%\subsubsection{\rqBatchInference}
\subsubsection{RQ3}

For each combination of 
\begin{inparaenum}[\it a)]
\item inference engine, \item  GPU model, \item  number of GPUs, and  \item \llm{}, 
\end{inparaenum}
we plot the evolution \throughput{} with the batch size. 
Then, for each combination, we test the null hypothesis \emph{$H_0$: There is no a significant difference in the throughput values based on the batch size} using Analysis of Variance test (ANOVA) 

%\subsubsection{\rqOlderHardare}
\subsubsection{RQ4}

To compare the \throughput{} obtained with a \nvidia{} A100 with that one obtained with a \nvidia{} V100, we  first compare the \throughput{} from pairs of trials, one from A100, the other from V100, but sharing the other hyperparameters (\llm{}, batch size, number of GPUs).
%This comparison is done using Eq.\ref{eq:improvementhard}:
%\begin{equation}
%\label{eq:improvementhard}
%\text{I$_{it}$} = \left( \frac{\text{throughput$_{it}$ A100} - \text{throughput$_{it}$  V100}}%{\text{throughput$_{it}$ on V100}} \right) \times 100
%\end{equation}
Moreover, we compute, for each inference engine, the Wilcoxon Signed-Rank Test to test the null hypothesis \emph{$H_0$:~The difference in \throughput{} between Nvidia A100 and v100 is not significant}.

%\subsubsection{\rqHPO}
\subsubsection{RQ5}
\label{sec:protocol:hpo}

We evaluate the impact of \hpo{} in two experiments: 
\begin{inparaenum}[\it 1)]
\item Upgrading and
    \item Downgrading  GPU. 
\end{inparaenum}
Both consider two GPUs models, \emph{previous} (P) and \emph{new} (N).
For studying the \emph{Upgrading}, P becomes the \nvidia{} V100 model, and N is \nvidia{} A100.
For \emph{Downgrading}, P is A100 and N is V100.
%Then, both experiment shares the same steps, with the difference on the values of N and P.
Each experiment is as follows.
For each pair \llm{} and inference engine, we  search the set of hyperparameters $HP_{orig}$ that maximize \throughput{} on P.
Then, we measure \throughput{} of using $HP_{orig}$ on N.
After that, we apply \hpo{} on N to find new hyperparameters $HP_{best}$ that maximize \throughput{} on N.
Finally, we compare \throughput{} from $HP_{best}$ and $HP_{orig}$  to measure the impact of \hfinference{}.

%subsubsection{Hyperopt \cite{Bergstra2015Hyperop}} is a distributed asynchronous hyper-parameter optimization framework, based on the Tree-of-Parzen-Estimators (TPE) algorithm \cite{Bergstra2011AlgorithmsHyper}.
%TPE is a Bayesian optimization approach~\cite{Shahriari2015}, based on \emph{Sequential Model-Based Optimization}~\cite{Hutter2011smbo}), which builds a probability model $p(score|hyperparameter)$ (aka the ``surrogate'' function) of the objective fitness function and uses it to select the most promising hyperparameters to evaluate in the objective function. 

To apply \hpo on inference engines, we implement \approachhpo{} (\underline{Inf}erence hyper\underline{P}arameter \underline{Op}timization).
Internally,  \approachhpo{}  uses Hyperopt~\cite{pmlr-v28-bergstra13Hyperopt}, a state-of-the-art \hpo{}, used  hyperoptimizing other software engineering tasks~\cite{lustosa2023optimizingSNEAK,martinez2023DAT}.
\approachhpo{} receives as input: 
\begin{inparaenum}[\it 1)]
\item a list of queries to pass to a LLM (in our experiment, queries for code-completion from the \emph{Single-line infilling} benchmark),
\item a specification of the hyperparameter space. In this experiment, we consider two hyperparameters: number of GPUs and batch size.
\item An \emph{objective/fitness function}: In this study, the function used by \approachhpo{} aims at maximizing inference \throughput{} on the mentioned queries.
\end{inparaenum}
\approachhpo{} produces as output a set of values for the hyperparameters that maximize the objective.
We hyperoptimize \vLLM{} and \hfinference{} on each \llm{}. 
Given the stochastic nature of Hyperopt, we conduct each experiment 10 times and report the average results.

\section{Results}
\label{sec:results}

%%%%%%% RQ 1

\subsection{\rqshapespace}

\input{figshapes}

Figure \ref{fig:landscapes} shows six plots, each corresponding to an LLM and an inference engine. In Appendix~\cite{appendix} we include all of them, including interactive versions that facilitate the visualization and analysis.
Two axes represent the hyperparameter space: batch size and number of GPUs, while the third one -the vertical one- corresponds to the throughput.
%These plots represent \todo{common} shapes that we observed between all plots generated.
This analysis does not aim to perform a rigorous classification of these shapes, but to provide evidence though these examples, that the throughput landscapes are not uniform, thus hyperparameter optimization would be useful. 

Figure \ref{fig:plot:space:a100_autohf_Bloom-1b7} shows the Bloom-1b7 landscape using \hfinference{}.
It has two peaks where the throughput is much higher ($\approx94.5$) than in other parts of the landscape (between 60 and 70 t/s): one uses one GPU and another uses eight GPUs, both with batch size 64.
Increasing that increasing the batch size above 64 produces memory errors.
%Reducing the batch size (from 64 up to 2) or using 2 or 4 GPUs significantly reduce the throughput (between 60 and 70 t/s).

The second case, Fig. \ref{fig:plot:space:a100_autohf_Codellama-7b-hf}, shows the landscape for CodeLlama and \hfinference{}. 
It presents different peaks, most notably:
\begin{inparaenum}[\it 1)]
\item  one single GPU and and batch size 16,
\item two GPUs and batch size 64, and
\item four GPUs and batch size 128.
\end{inparaenum}
This shows that for CodeLlama when one increases the GPUs, the batch size should be also adjusted.
This case also shows the challenging of determining optimal hyperparameter values:
Codellama-7B using one single GPU can infer batches up to size 16. Putting values of 32 or greater produces memory errors.

The third case, Fig. \ref{fig:plot:space:a100_autohf_Mistral-7b} corresponds to Mistral-7b  and \hfinference{}.
It has a peak at one GPU and batch size 32. Greater batches with one GPU cause memory errors. 
Beyond this peak, the landscape presents a descending inclined plane as the batch enlarges. 
The landscape also exhibits high values of throughput (similar to the mentioned peak) with four GPU and small batches (4 and 8).
With 4 GPUs, increasing the batch size decreases the throughput.

The fourth case, 
Fig. \ref{fig:plot:space:vLLM_Gemma-2b} corresponds to Gemma-2 and \vLLM{}.
We observe that:
\begin{inparaenum}[\it 1)]
  \item There is a significant improvement of throughput when using 4 or 8 GPUS compared to using 1 or 2 GPUs.
 \item The landscape has a \emph{plateau}: a vast zone where the values are relatively constant, do not change significantly throughout the plateau, and represent a high-throughput region. We can observe it in yellow.
This plateau is located in the region bounded by 4 to 8 GPUs and batch sizes ranging from 128 to 1024.
  \item The throughput given with eight GPUs is slightly higher than that with four.
 \item The highest throughput is with a batch size equal to 258: increasing the size slightly impacts throughput. However, the reduction is more evident when the batch size is smaller than 258.
 \end{inparaenum}

%The fifth case, Fig.\ref{fig:plot:space:autohf_Codestral-22b} corresponds to CodeStral-22b and \hfinference{}.
%It produced results with 4 and 8 GPUs, but not for \todo{1 or 2}.
%The plot shows a flat region that maximizes throughput for batch sizes between 4 and 16.
%Neverteless, given a particular batch size, the throughput given with four GPUs is slightly higher than those with eight. 

The fifth case, Fig. \ref{fig:plot:space:a100_autohf_Starcoder} corresponds to Starcoder-15b and \hfinference{}.
It has a peak with one GPU and batch size 32 and increasing the batch size to a value just greater than the peak produces a memory error.
However, this landscape has two main differences with respect to the two previous examples.
First, it has a second peak when using two GPUs and a larger batch size (128).
Secondly,
the throughput landscape shows a downward slope, indicating that throughput decreases as the number of GPUs increases.

{\bf\emph{Hardware specifics.}}
The previous landscapes resulted from the execution of inference on a computing node equipped with \nvidia A100. %, one of the most widely used hardware to train the evaluated model.
We also inspect the landscape resulted from the  V100.
Even in the majority of cases the landscapes from both models produces similar shapes (A100 shapes have higher peaks, i.e. higher throughput, because it is a more recent and advanced model), others exhibit more noticeable differences.
For example, Fig.~\ref{fig:plot:space:tesla_autohf_Starcoder} shows the execution of StarCoder-15b in a V100 (the Fig at its left shows the landscape of this model from A100).
We observe that the shape is different: there is not a peak but a thin and long plateau region bounded by 2 and 8 GPUs batch sizes 8 and 16.
This means that, a \llm{} is deployed in a another GPU model,
it might be necessary to adjust the hyperparameters in order to maximize the throughput.
In RQ5 we study this.

%\medskip
\begin{graybox}
%Visual inspection of the throughput landscape from  inference using \hfinference{} and \vLLM{} shaped by different hyperparameter values shows that the throughput is sensitive to hyperparameter and GPU hardware. 
Visual inspection of the throughput landscape from  inference using \hfinference{} and \vLLM{} shows irregular landscapes, some with marked peaks.
Consequently, to ensure \throughput{} maximization for a particular inference engine and hardware would require an inspection of the landscape.
\end{graybox}
%\medskip

%In the rest of this paper,  we perform a detailed, quantitative study, focusing on specific \throughput{} values to derive precise insights and statistical significance.

%Scaling Efficiency
%Communication overhead: With multiple GPUs, inter-GPU communication can become a bottleneck, especially as you scale up.

%Memory constraints: Larger batch sizes require more GPU memory, which can be a limiting factor.
%Scaling efficiency: As you increase the number of GPUs, you'll want to monitor how well the throughput scales (linear scaling is ideal but rarely achieved in practice).
%Communication overhead: With multiple GPUs, inter-GPU communication can become a bottleneck, especially as you scale up.
%Hardware specifics: Different GPU models and interconnect technologies can significantly impact performance.

%%%%%%%%%%%%%%%% ONLINE:

\input{tables/tabHFvLLMOnline}
\input{figOnlineHFGPUs}

\subsection{\rqonlineinferenceHFGPU}

\autoref{tab:minMaxMetricAutoHFVLLMOnlineInferece}  and \autoref{fig:onlineGPUsAutoHFvLLM} show the throughput of online inference performed by \hfinference{} and \vLLM{} when scaling \nvidia{} A100 GPUs.
The table shows: 
\begin{inparaenum}
    \item for \hfinference{}, the number of GPUs that maximizes the \throughput{} for each model (column Best),
    \item for \vLLM{}, in column Same, the \throughput{} using that number of GPUs (for a fair comparison),
    \item for \vLLM,  the number of GPUs that maximizes the \throughput{} for each model (column Best).
\end{inparaenum}
Tables showing the \throughput{} per GPU, and the results from V100 can be found in our appendix~\cite{appendix}.

%Let us start the analysis by analyzing the figures. 
We observe that the scaling hardware when using \hf{} has a different effect than when using \vLLM{}. All findings also apply for Nvidia V100.
%For this reason, we study them separately.

\subsubsection{\bf \vLLM{}}
From \autoref{fig:gpus_evol_a100_vLLM_throughput}  we identify two primary observations.
\begin{inparaenum}
    \item[] First, scaling  from 1 to 2 GPUs using \vLLM{} has a strong (negative) impact on \throughput{} of some models (e.g., Microsoft Phi-1), even for others (such as CodeLlama-13b) has a moderate (positive) impact.
    \item[] Second, scaling from 2 to 4, and then from 4 to 8 GPUs has less impact on \throughput{}: the plot shows that the \throughput{} of most of the model is bounded between 60 and 80 \ts{}. 
\end{inparaenum}

We also observe:
\begin{inparaenum}[\it a)]
    \item with one GPU, \throughput{} presents a large variability in the models: it goes from 32.5 to 171.5 \ts{}.
    This variability is reduced after scaling.
    \item 8 models with $\le$ 3 billion parameters  maximize \throughput{} using a single GPU,  
    \item when scaling to two GPUs, there are two effects: the first one is that \throughput{} dramatically drops (which is the case for the ``smaller'' models previously mentioned); the second one is the \throughput{} slinging vary positively in most of  models with 13-15 billion parameters.
    \item  Models with 7-13 billion parameters maximize \throughput{} using 4 GPUs. %Thus, \todo{scaling up to 4 GPS has a negative impact for these models}
    \item  Models with more than 15 billions parameters typicaly maximize \throughput{} with 8 GPUS.
    %\item The largest models we evaluated (e.g., CodeLlama-34b, WizardCoder-33b) maximize \ts{} with 8 GPUs (in total, 6 models, see column \texttt{\vLLM{}-Best}). 
    %However, reducing the scale does not cause substantial losses \ts{} 
    %(e.g., the \throughput{} from StarCoder-15b  is 57 and 58.4 for 4 and 8 GPUs respectively).
   % In this case, the developer and/or manager and/or infrastructure leader need to assess the value of scaling up – and the additional cost of resources – to achieve those extra throughput points and consider how this could lead to increased business value.

\end{inparaenum}

\paragraph{\bf Statistical test for \vLLM{}}
%We performed a hypothesis test to determine whether the throughput values differ significantly according to the number of GPU uses.
%We perform an analysis of variance (ANOVA) as we have multiple groups, each for a number of GPUs used on inference (1, 2, 4 and 8).
Taking into account all GPUs, the ANOVA test rejects the null hypothesis $H_0$ at the significant level $\alpha =  0.05$ (F-statistic: 5.44, p-value: 0.002) which indicates that there is a statistically significant difference in throughput across different numbers of GPUs for both A100 and V100 GPUs.
We suspect that this significant difference is dragged by one of the analyzed groups, GPU 1, which maximizes \throughput{} for some \llm.
For that, we performed two additional tests. First, the re-execution of the ANOVA test by just considering trials with 2, 4 and 8 GPUs  failed to reject $H_0$.
Secondly,  Tukey HSD test (Honestly Significant Difference) reveals significant differences in performance metrics between using GPU 1 and each of the other groups (2, 4, and 8). However, no significant differences were found between GPUs 2, 4, and 8. 
%which provides a pairwise comparison of the means of different groups to determine if their differences are statistically significant.
%The test results reveal significant differences in performance metrics between GPU 1 and GPUs 2, 4, and 8. However, no significant differences were found among GPUs 2, 4, and 8. 

\subsubsection{\bf \hfinference{}}
From  \autoref{fig:gpus_evol_a100_autohf_throughput}, we observe two major differences with respect to \vLLM{}:
\begin{inparaenum}[\it 1)]
    \item the \throughput{} are lower than using \vLLM{}, independently of the number of GPUs used;
    \item when scaling, there is no dominant pattern evolution of \throughput{} as for \vLLM{}.
\end{inparaenum}
Moreover:
\begin{inparaenum}[\it a)]
\item Scaling GPU results in a very slight decrease of \throughput{} from  models with ~7b parameters  (Bloom-7B1, CodeLlama-7b, Mistral-7b).
\item The decline is more pronounced in StarCoder2-3b. 
\item Scaling cause down and up (e.g., Bloom-1b7, Bloom3-b).
\item Scaling cause up and down (e.g., Phi-1, Phi-2).
\end{inparaenum}
In conclusion, using the  \hfinference{}, scaling generally does not improve throughput for most LLMs (with few exceptions such as  Phi-1 and Gemma-7). 

\paragraph{\bf Statistical test for \hfinference{}}

Based on the results of the ANOVA test, we cannot reject the null hypothesis ($H_0$) at the 0.05 significance level (F-statistic: 0.37, p-value: 0.77). This means that there is not enough statistical evidence to suggest that the means of the groups -number of GPUs- are significantly different, which can be translated as scaling up GPUs does not produce significant changes in \throughput{}.

\subsubsection{\bf Comparison between Inference Engines}
Table \ref{tab:minMaxMetricAutoHFVLLMOnlineInferece} shows the \throughput{} for each model (columns \texttt{Thp}) and the improvement in \throughput{}.
Taking into account the number of GPUs that led \hfinference{} to maximize \throughput{} (column \texttt{\vLLM{} Same}), \vLLM{} achieves a mean (median) improvement of 4.3$X$  (6.47$X$).
However, for 12 LLMs, \vLLM{} maximizes \throughput{} by using a different number of GPUs (column \texttt{\vLLM{} Best}), 
achieving a mean (median) improvement of 7.26$X$ (4.80$X$) with respect to \hfinference{}.
In nine of these 12 cases, \vLLM{} achieves the gain by using more GPUs and, in the remaining three cases, fewer GPUs.
This shows that both \vLLM{} and \hfinference{} need to be optimized separately, at least on the number of GPUs, to increase \throughput{}.

The reason behind the difference in \throughput is that \vLLM{} was designed to achieve fast inference.
For that, \vLLM{} introduces key improvements. We briefly discuss two of them.
First, \vLLM{}  uses \emph{PagedAttention}~\cite{kwon2023efficient},  a new attention algorithm that manages the attention keys and values from the Transformer architecture \cite{vaswani2017attention}.
In autoregressive decoding, all the input tokens to the LLM produce their attention key and value tensors, and these tensors are kept in GPU memory to generate the next tokens. 
This has several problems for LLMs, such as the large amount of memory required and memory waste~\cite{kwon2023vllmBlog}. 
PagedAttention proposes a solution inspired by the idea of virtual memory and paging in operating systems: it allows storing continuous keys and values in noncontiguous memory space.
The second improvement is \emph{Continuous batching}, initially proposed by~\cite{Orca2022Batching}.
During inference, instead of waiting until every sequence in a batch has completed generation (as traditional static batching does) this approach, once \emph{a sequence} in a batch has completed generation, a new sequence is inserted in its place. This yields a higher resource utilization than static batching \cite{vllmcontinousbatching} and thus increases \throughput{}.

% https://huggingface.co/docs/transformers/perf_infer_gpu_one
%https://huggingface.co/blog/bloom-inference-optimization

%\medskip
\begin{graybox}
{\bf \underline{Answer to RQ2:}} 
For \vLLM{}, GPU scalability benefits larger LLMs ($\approx$13 billion parameters or more), maximizing \throughput{} with the use of eight GPUs. However, scaling up GPUs penalizes smaller models ($\approx$ 3 billion parameters or fewer), decreasing their \throughput{}. For these smaller models, the optimal configuration is using 1 or 2 GPUs.
In contrast, using \hfinference{}, scaling hardware does not significantly increase \throughput{} for most of the LLMs studied. 
Additionally, on the same hardware, \vLLM{} achieves an average \throughput{} that is 6.47 times higher than that of \hfinference{}.
\end{graybox}
%\medskip

%\medskip
\begin{graybox}
{\bf \underline{Implications:}}
 \hfinference{}{} may conduct to under-optimal results when the node has more than one GPU (since it uses all GPUs).
Furthermore, we recommend hyperoptimizing the number of GPUs to be used during inference. 
%and explicitly pass this parameter rather than, for instance, using the \texttt{auto} on hyperparameter \texttt{device\_map}.
\end{graybox}
%\medskip

%%%%%%% RQ 3
\subsection{\rqBatchInference}

\input{figBatchesAll}

We study how the \throughput{} varies according to the batch size by using a particular number of GPUs.

\subsubsection{\bf \hfinference{}}

\autoref{fig:batchesA100} shows the \throughput{} using 1, 2, 4 and 8 \nvidia{} A100 GPUs  and \hfinference{.}%\footnote{To better visualize the trends, we remove our outlier (StarCoder-15b), which is shown in \autoref{fig:RQ3_bs_throughput_a100_autohf_all}}
We observe:
\begin{inparaenum}[\it a)]
    \item The four plots show similar trend.
     \item Scaling on hardware (more GPUs) allow to have larger batches. 
The reason is that there is more memory on the GPU to allocate bigger batches and thus compute more inputs in parallel.
  \item the maximum \throughput{} is usually reached with the largest batch size that correctly works. For example, Phi-1 achieves the maximum \throughput{} with a batch size 64, larger batches produce memory errors. 
  %   \item for some \llm, increasing the batch size has a impact on the \throughput{}, for example, StarCoder-15.
     
\end{inparaenum}

\paragraph{\bf Statistical test for \hfinference{}}

We compute the ANOVA test for each group of GPU configurations (1, 2, 4, and 8).
For both  \nvidia{} A100 and V100, we reject the null hypothesis $H_0$ for 1, 2 and 4 GPUs, at the significant level $\alpha=0.05$. All p-values are shown in our appendix.
In contrast, we cannot reject $H_0$ when the trial has eight GPUs.
%We hypothesize that the increase of GPUs to eight also entails having more memory to process larger batch sizes and, consequently, .
%\todo{here improve}

\input{figBatchSize}

\subsubsection{\bf \vLLM{}}

\autoref{fig:RQ3_bs_throughput_a100_autohf} shows the \throughput{} obtaining with different batch sizes and using two GPUs.
The plots for one, four, and eight GPUs exhibit a similar trend and are available in our appendix \cite{appendix}.
From these plots, we observe that:
\begin{inparaenum}[\it 1)]
    \item Throughput generally increases with batch size across most models.
    \item The improvement is more noticeable for smaller to medium batches (up to 128).
   \item Most models achieve their highest throughput around batch sizes of 64 to 256. 
   Beyond this point, throughput tends to plateau, indicating that increasing batch size further does not significantly improve throughput.
   \item For some models, such as Starcoder-15b and Gemma-2b, after the plateau, the \throughput{} shows a slight, but not significant, decrease.
   \item  Moreover, for Bloom-1b7 (as well as for other models using 4 or 8 GPUS) the \throughput{} shows a slight decrease followed by a subsequent increase.
\end{inparaenum}
All these points mean that optimizing the batch size is crucial to maximize \throughput{} in \llm{}. 
when using \vLLM{} inference.

\paragraph{\bf{Statistical test}}

For both types of hardware, \nvidia{} A100 and V100, and for each number of GPUs, we reject the null hypothesis $H_0$ at the significant level $\alpha=0.05$ (p-values listed in the appendix).

\subsubsection{\bf{Comparison \hfinference{} and \vLLM{}}}

\autoref{fig:batchsizes2GPU} shows the \throughput{}  from \hfinference{} and \vLLM{} both using two \nvidia{} A100 GPUs)
We observe that:
\begin{inparaenum}[\it a)]
    \item with batch size 1, in both engines the \throughput{} is below to 100 \ts{}.
    \item However,  \throughput{}  from \vLLM{} increases rapidly with batch size. For \hfinference{} the increase is moderate for some models (e.g., Phi-1, Bloom-1b7) or null for others (CodeLlama-7b).  
    \item Starcoder-2-3b with \hfinference{} outperforms the other models and has a huge increase from batch size 4.
     \item Since batch size 4, the \throughput{}  from \vLLM is between 100 and almost 800 \ts{}. In contrast, for \hfinference{}, with the exception of the Starcoder-2-3b mentioned above, \throughput{} never passes 100 \ts{}.
\end{inparaenum}
These findings also apply for the inference using 1, 4 and 8 GPUS and GPU \nvidia{} V100.

\paragraph{\bf{Statistical test}}

We apply Wilcoxon Signed-Rank Test to test the null hytothesis $H_0$: 
\emph{The difference in throughput between \vLLM{} and \hfinference{} is not significant.}
For all number and model of GPUs, the test rejects $H_0$ at the 0.05 significance level (all p-values are in our appendix).
Moreover, we measure the effect size employing Cliff's delta ($\delta$) in order to quantify the magnitude of differences between the \throughput{} from \vLLM{} and \hfinference{}.
Across all the number and models of GPUs evaluated, the mean $\delta$ is 0.8762.
This indicates a very large effect size, suggesting a significant difference between the two groups,  with \vLLM{}  scoring higher than \hfinference{}.

%\medskip
\begin{graybox}
{\bf \underline{Answer to RQ3:}} 
In \hfinference{}, increasing the batch size leads to slight or moderate increases in \throughput{} for some models (e.g., CodeLlama-7b) and steep increases for others (e.g., Bloom-1b7).
Increasing the size of the batch that maximize the \throughput{}  often results in out-of-memory errors.
Scaling up GPUs allows \hfinference{} to process larger batches.
Conversely, increasing \vLLM{} significantly impacts small to medium batches (fewer than 64). For larger batches, changes in \throughput{} are not substantial and do not cause out-of-memory. 
%\vLLM{} has significantly higher \throughput{} than \hfinference{} in the two hardware devices tested.
Overall, \vLLM{} achieves significantly higher \throughput{} than \hfinference{} on the two hardware devices tested.
\end{graybox}
%\medskip

%\todo{Bad things of increasing the batch size in bt}

%\medskip
\begin{graybox}
{\bf \underline{Implications:}} 
Optimizing batch size is crucial to maximize \throughput{} in \llm{}. 
In \hfinference{}, determining the optimal batch size and number of GPUs to achieve maximum \throughput{} is challenging because is bonded by the GPU memory. %due to the limitations imposed by \nvidia{} GPU memory.
%ecause it is bounded by the \nvidia{} GPU memory.
This insight is valuable for hyperparameter optimization, especially in scenarios where throughput is a critical performance metric.
\end{graybox}
%\medskip

\subsection{\rqOlderHardare}

%\input{tables/tabHardwareComparison}

%We compute the differences of \throughput{} between \nvidia{} A100 and V100. 
For \vLLM{}, the increase of \throughput{} using A100 compared to V100 is on average 83.5\% (median 77.6). 
The models that exhibit the largest difference are Gemma-7b, Codellama-13b, 34b and 70b, Wizardcoder-15b, all with 100\% or greater improvement by using A100.
%We did not detect a correlation between model size and improvements, e.g. WizardCoder-33b shows a 57.4\% of improvement, improvement similar to a smaller model, Phi-1\_5 with 51.4\%, but lower than a model with similar number of parameters Codellama-34b with 114.1\%.
Appendix \cite{appendix} includes a detailed table with the improvement for each model and hardware. 

We compute Wilcoxon Signed-Rank Test to test the null hypothesis $H_0$: 
\emph{The difference in throughput between \nvidia{} A100 and V100 is not significant} for \vLLM{}. We reject $H_0$ at the significant level $\alpha = 0.05$  (p-value = 2.01).
We measure the effect size to quantify the differences of \throughput{} between A100 and V100 using Cliff's delta~\cite{cliff1993dominance}. A delta value of 0.406 suggests a moderate effect size, indicating a moderate difference between A100 and V100 using \vLLM{}.

For \hfinference{}, the increase of  \throughput{}  using A100 is lower on average, 47\% (median 23.3\%).
Notally, the increase is greater  for the largest models such as WizardCoder-33b (535.6\%) and Codellama-70b-hf (276\%), but smaller for the small models such as Gemma-2b (1.9\%),  Phi-2 (21.8\%).
We reject $H_0$  for \hfinference{} at the significant level $\alpha = 0.05$ (p-value = 3.88).
The measurement of effect size using Cliff's delta is 0.1333, which indicates a small but noticeable effect accross   \throughput{}  from A100 and V100.

\begin{graybox}
{\bf \underline{Answer to RQ4:}} 
Upgrading (or downgrading) the GPU model has a significant impact on the throughput of \vLLM{}, with an improvement of 83.5\% on average) and on \hfinference{} with an improvement of 47\% on average.

\end{graybox}

\begin{graybox}
{\bf \underline{Implications:}} 
Developers of AI-based systems need to analyze the following trade-off, either:
\begin{inparaenum}[\it a)]
    \item  {\underline {Maximizing Throughput}}: Opt for the \nvidia A100 to achieve maximum throughput. This choice involves higher costs due to the premium price of state-of-the-art hardware; or % driven by high demand from the AI boom.
\item  {\underline {Reducing Operational Costs}}: Choose \emph{older} hardware like \nvidia{} V100, which might result in lower throughput but eventually reduce operational expenses. This option is particularly advantageous for LLMs where the difference in throughput between A100 and V100 is modest (e.g., \hfinference{} and small LLMs such as Microsoft-Phi).
\end{inparaenum}
\end{graybox}

\subsection{\rqHPO}
\label{sec:rqhpo}

\input{figHPO}

Fig \ref{fig:violinhpo} shows the distribution of improvement in \throughput{} given by the new hyperparameter values found by our tool \approachhpo{},  in the context of GPU upgrade (from \nvidia{} V100 to A100) and downgrade (from  A100 to V100).
For upgrade, optimization produces \hfinference{} a median increase (avg) in \throughput{} of 6.4\% (9.16\%).
For some models, e.g., Starcoder-15b the improvement is higher (24\%), while for others three there is no improvement (i.e., the best configuration, in terms of \throughput{}, of \hfinference{} on V100 is also the best on A100).
For downgrade, the median improvements are 5.61\% (average 13.7\%).
For \vLLM{} the improvements are lower, in both upgrading (median 3.3\%) and downgrading (median 6.9\%).
The reason is that, as shown in \autoref{fig:batchsizes2GPU}, optimizing the evaluated hyperparameters in \vLLM{} has a lesser impact on \throughput{} than in \hfinference{}.

\begin{graybox}
{\bf \underline{Answer to RQ5:}} 
Our results suggest that applying hyperparameter optimization during changes in the inference infrastructure —such as upgrades or downgrades of GPU devices— can yield improvements in throughput. On average, \hpo{} enhances throughput from \hfinference{} by 9.16\% during GPU upgrades and 13.7\% during GPU downgrades.
\end{graybox}

\begin{graybox}
{\bf \underline{Implications:}} 
Applying \hpo{} during hardware upgrades (e.g., GPU) is crucial to maximize the benefits provided by the new hardware. Applying \hpo{} during hardware downgrades is essential because the transition from a newer model (A100) to an older one (V100) typically results in a decrease in \throughput{}. 
This study demonstrates that \hpo{} can help to mitigate these performance losses.
%\todo{re-tuning}

\end{graybox}

\section{Discussion}
\label{sec:discussion}

\subsection{{Deployment of LLMs on available hardware}}

%\todo{REASONS for scalling decrease the results}
%\todo{It first  use the maximum space available on the GPU(s)
%if we still need space, we store the remaining weights on the CPU}
%https://huggingface.co/docs/accelerate/v0.32.0/en/concept_guides/big_model_inference

Both \vLLM{} and \hfinference{} inference engines are able to deploy the largest \llm{} using the CPU and just one GPU.
The reason is that these engines provide specially  mechanisms, such as HF Accelerate \cite{accelerate} activated via the hyperparameter   \texttt{device\_map="auto"}, to  determine automatically where to put each layer of the model depending on the available resources.
In the case of Accelerate, it first uses the maximum space available on the GPU(s), and when it still needs space, it stores the remaining weights on the CPU, RAM, and finally on the hard drive as memory-mapped tensors.
For example,  Starcoder2-3b and CodeLlama-7b fully fit in one GPU (either A100 or V100).
From Table \ref{tab:minMaxMetricAutoHFVLLMOnlineInferece} we observe using just 1 GPU maximizes their \throughput{}.
For them, if one uses \hfinference{} with \texttt{device\_map="auto"} on a infrastructure with more than one A100 GPU, the \throughput{} would be negatively affected.
The other models can perform the inference with 1 GPU, but also needs to store weights on CPUs and eventually on disk (e.g., Mixtral), damaging the \throughput{}.
%This is a example that having more GPUs negatively impacts on \hfinference{} when using  \texttt{device\_map="auto"}.

The amount of GPU memory has an impact.
For example, \hfinference{} achieves to store all CodeLlama-70b weights in eight A100s, each with 40Gb. 
However,  it cannot do that on the eight V100 with 32GB, which requires the storage of some parameters in the CPU. This has an impact on \throughput{} (1.79 vs 0.21), beyond the technological improvements introduced by the \nvidia{} A100.

\subsection{Correlation between \throughput{}, HumanEval score and number of parameters}

Table \ref{tab:minMaxMetricAutoHFVLLMOnlineInferece} shows that, for both \hfinference{} and \vLLM{}, those that maximize \throughput are the models with smaller number of parameters, such as Bloom-1b7, Gemma-2, and Starcoder2-3b.
The Pearson correlation between parameters and \throughput (columns \texttt{\hfinference{} Best}, \texttt{vLLM Same} and \texttt{vLLM Best}) returns coefficients between -0.623 and -0.67, suggesting a noticeable inverse relationship between throughput and the number of parameters (p-values between 0.001 and 0.003). 

%We also analyze the trade-off between model size and \throughput{}.
The largest models generally achieve the highest performance scores in HumanEval, as shown in \autoref{tab:model_data}.
The correlation coefficient between HumanEval and the parameters is 0.527 (p-value 0.017), indicating a moderate positive relationship. This suggests that models with more parameters tend to have higher HumanEval scores. 

We also compare the correlation between \throughput{} and HumanEval.
For \hfinference{} the coefficient is $-0.58$ (p-value 0.0072). As throughput increases, HumanEval tends to decrease.
For \vLLM{}, considering the `best' \throughput{} (\autoref{tab:minMaxMetricAutoHFVLLMOnlineInferece}) the coefficient is -0.35 (p-value 0.122): As throughput increases, HumanEval tends to decrease slightly. At the significant level $\alpha=0.05$, the correlation is not significant.
These results suggest that \vLLM{} achieves a more favorable trade-off between performance and throughput compared to \hfinference.

\subsection{Threats to Validity}

It could be the risk that HumanEval is not representative of the code completion task. Neverteless, we consider it because it has been used on similar tasks such as infilling \cite{fried2022incoder,bavarian2022efficient}.  

There are other inference engines that could eventually achieve state-of-the-art throughput, such as TensorRT-LLM~\cite{tensorRTLLM}.
We choose \vLLM{} over other efficient engines because it first introduced a novel mechanism (PagedAttention~\cite{kwon2023efficient}), later adopted by other engines such as TensorRT-LLM.

%other parameters that affects \throughput{} that we do not focus.
%We focus on two

\section{Related work}
\label{sec:relatedwork}

%\subsection{Inference engines}

%\cite{kwon2023vllmBlog} comparison HF and vLLM.
%\cite{yu2023characterizing} 
%\subsection{HP0}

Wang et al. \cite{wang23HOPLLM} presented EcoOptiGen, a hyperparameter optimization for LLM inference.
Their and our work have substantial differences.
They hyperoptimize the inference of closed LLM from OpenAI, accessible through an API for optimizing accuracy.
In contrast, our work optimizes on throughput by exploring parameters such as number of GPUs, batch sizes, which are not accessible via the OpenAI API. %, and we focus on optimizing open LLMs. 

%Our work focus optimize hyperparameter in order to avoid issues such as out of memory during inference. 
Previous work has study inference issues.  Gao et al. \cite{Gao2024GPUlowIcse}  studied low GPU utilization issues from jobs submitted to Platform-X, a Microsoft internal deep learning platform.
They classified 706 issues discovered across 400 deep learning jobs, and found that the 25.64\% of the issues were related to ``Improper Batch Size'' and 
the 3.12\% are related to ``Insufficient GPU Memory".  
Zhang et al. \cite{Zhang2020FailuresJobs} conducted a similar study another deep learning platform in Microsoft (Philly) and found that 8.8\% of the issues were related to GPU out-of-memory.
These results show that \approachhpo{} (after changing the fitness function) could be useful to avoid these types of issues.
Motivated by the out of memory failures, tools such as DNNMem~\cite{Gao2020EstimatingGPUconsumption} and DNNPerf~\cite{Gao2023RuntimePerformancePrediction}  were proposed for predicting  GPU memory consumption and performance of deep learning models. 
Similarly, Cao et al. \cite{cao2022understanding} studied performance problems of TensorFLow and Keras frameworks from StackOverflow posts.
They found 12 performance problems related to ``Improper Hyper Parameter''.  
In future work, \approachhpo{}  could be adapted to try to fix them.
%Note that none of these worked focused on large language models.

Previous work has also focused on studying training and/or inference configuration via hyperparameters.
McCandlish et al. \cite{mccandlish2018empirical} predicted the largest useful batch size for neural network training, addressing the trade-off between training speed and computational efficiency. In our paper, we focus on inference.
Gao et al.~\cite{gao2021resource} proposed DnnSAT, a resource-guided approach for deep learning models to help existing AutoML tools ~\cite{hutter2019automated} (such as Hyperparameter Optimization)  efficiently reduce the configuration space.
%They  evaluate the reduction effectiveness of DnnSAT on two models (not LLMs)  achieving of a reduction of the configuration space up to 53.4\% and 24.1\% respectively.  
DnnSAT could be integrated into our tool \approachhpo{} to reduce the number of configurations.
%Islam et al.~\cite{islam2019comprehensive}  study of bugs in the usage of deep learning libraries inclucuding TensorFlow and Torch by analyzing StackOverflow post and commits. They found that Incorrect Model Parameter is one of the most of the bugs happen in the Data Preparation stage of the deep learning pipeline.
%These kind of problems can be solved by some automated model and parameter recommendation tools.

%Xhou et al. \cite{zhou2024survey} survey on efficient inference  for LLMs, which  targets different aspects, from providing primary causes of
%the inefficient LLM inference to  defining comprehensive taxonomies of, for instance, inference engines.
%In particular, they analyze features from inference engines including the two we consider, \hfinference and \vLLM{}
%They also conducted experiments such as the evaluation of the speed-ups achieved
%by employing uantization techniques on two models LLaMA-2-7B and 13B using two inference frameworks, TensorRT-LLM\cite{}  and LMDeploy\cite{}. %Recent advances on inference are also discussed in \cite{li2024llm}
%They used a single single NVIDIA A100 GPU

%%%%%%%%%%%%%%%%%%%%%%%%%%%%%%%%%%%

Previous work has conducted related empirical work on LLM inferences.
Samsi et al. \cite{Samsi2023BenchmarkingInferenceLlama} benchmarked the energy consumed by Llama models: 7b, 13b and 65B.
Beyond the research goals, there are some differences with our work, notably the model evaluated (3 vs. 20) and the inference libraries (Llama scripts~\cite{llamaScripts}, based on Pytorch and FairScale~\cite{FairScale2021}, vs \hfinference{} and \vLLM{}).
Xhou et al. \cite{zhou2024survey} evaluated the speed-ups achieved
by employing quantization techniques on two models LLaMA-2-7B and 13B. %using two inference frameworks, TensorRT-LLM~\cite{tensorRTLLM}  and LMDeploy~\cite{2023lmdeploy}.
Argerich et al. \cite{Argerich2024Inference} presented EnergyMeter, a software-based tool to measure the energy consumption of different hardware
components. For evaluation, they considered LLMs with at most 7b parameters (we consider LLMs up to 70b) and use HuggingFace Transformers for inference.
Coignion et al.~\cite{coignion:hal-04525620} evaluated the code generated by 18 LLMs on the Leetcode benchmark. 
The main differences from our work, beyond the goal, are the LLMs evaluated (they consider smaller LLMs, up to 15b parameters) and the inference engine used (DeepSpeed vs \hfinference and \vLLM{).
Yarally et al. \cite{Yarally2023BatchingGreenAI}  studied how batch inference affects the energy consumption of computer vision tasks using Pytorch.
Alizadeh et al. \cite{alizadeh2024green} evaluated the energy consumption of deep learning runtime infrastructures such as ONNX, PyTorch and TensorFlow. 
%They also considered two batch sizes (1 and 64). 
Beyond differences in research goals, these two works do not focus on LLMs.

\section{Conclusions and Future work}
\label{sec:conclusion}
\label{sec:fw}

In this paper, we conduct a large-scale evaluation of the \throughput{} of 20 LLMs using two inference engines.
We inspect the \throughput landscape from the hyperparameter space and demonstrate that hyperparameter optimization can be applied to improve \throughput{}.

In future work, we plan to:
\begin{inparaenum}[\it 1) ]
\item Evaluate the performance of other inference engines such as those listed in the surveys~\cite{miao2023towards,zhou2024survey}.
\item Analyze the landscapes from other measures such as resource utilization (CPU, GPU)  or latency. 
\item Incorporate these measures into \approachhpo{}'s  optimization objective.
\item Evaluate other hyperparameter optimization approaches and space reduction techniques such as \cite{cao2022understanding} 
\item Study and optimize others hyperparameters that may affect \throughput{} and the eventual new measures.
\item Incorporate the specification of these hyperparameter spaces into \approachhpo{}.
\item Release a usable version of \approachhpo{} and present its architecture, design, implementation decisions, and usage in a demo/tool track paper.

\end{inparaenum}

\bibliographystyle{plain}
\bibliography{references.bib}

%\newpage
%\appendix
%\input{appendix}

\end{document}

%% file: tables/tabLLMInfo.tex
\begin{table}[t!]
\centering
\begin{tabular}{|l|r|r|}
\hline
\textbf{HuggingFace model id} & \textbf{\#Params} & \textbf{HumanEval} \\ 
\textbf{} & \textbf{(Billion)} & \textbf{(pass@1)} \\ \hline

\href{https://huggingface.co/bigscience/bloom-1b7}{bigscience/bloom-1b7} \cite{workshop2022bloom} & 1.7& 4.4   \\ \hline
\href{https://huggingface.co/bigscience/bloom-3b}{bigscience/bloom-3} \cite{workshop2022bloom} &  3 & 6.3   \\ \hline
\href{https://huggingface.co/bigscience/bloom-7b1}{bigscience/bloom-7b1} \cite{workshop2022bloom} & 7.1&  8.1   \\ \hline
\href{https://huggingface.co/codellama/CodeLlama-7b-hf}{facebook/CodeLlama-7b} \cite{roziere2023codellama}& 7 & 38.4 \\ \hline
\href{https://huggingface.co/codellama/CodeLlama-13b-hf}{facebook/CodeLlama-13b} \cite{roziere2023codellama}& 13 & 43.3 \\ \hline
\href{https://huggingface.co/codellama/CodeLlama-34b-hf}{facebook/CodeLlama-34b} \cite{roziere2023codellama}& 34 & 53.7 \\ \hline
\href{https://huggingface.co/codellama/CodeLlama-70b-hf}{facebook/CodeLlama-70b} & 70 & 57.3 \\ \hline
\href{https://huggingface.co/mistralai/Codestral-22B-v0.1}{mistralai/Codestral-22B} & 22 & 81.1 \\ \hline
\href{https://huggingface.co/mistralai/Mistral-7B-v0.1}{mistralai/Mistral-7B-v0.1} \cite{jiang2023mistral}& 7 & 30.5 \\ \hline
\href{https://huggingface.co/mistralai/Mixtral-8x7B-v0.1}{mistralai/Mixtral-8x7B-v0.1} \cite{jiang2024mixtral}& 46.7& 40.2 \\ \hline
\href{https://huggingface.co/google/gemma-2b}{gemma-ai/gemma-2b} \cite{team2024gemma}& 2 & 22.0 \\ \hline
\href{https://huggingface.co/google/gemma-7b}{gemma-ai/gemma-7b} \cite{team2024gemma}& 7 & 32.3 \\ \hline
\href{https://huggingface.co/microsoft/phi-1}{microsoft/phi-1} \cite{gunasekar2023textbooksPhi_1} & 1.3 & 50.6 \\ \hline
\href{https://huggingface.co/microsoft/phi-1_5}{microsoft/phi-1.5}  \cite{Li2023Phi1_5}& 1.3 & 	41.4 \\ \hline
\href{https://huggingface.co/microsoft/phi-2}{microsoft/phi-2} \cite{javaheripi2023phi2}& 2.7 & 49.4 \\ \hline
\href{https://huggingface.co/bigcode/starcoder}{bigcode/starcoder-15b} \cite{li2023starcoder}& 15.5 & 28.7\\ \hline
\href{https://huggingface.co/bigcode/starcoder2-3b}{bigcode/starcoder2-3b} \cite{lozhkov2024starcoder}& 3 & 31.7 \\ \hline
\href{https://huggingface.co/bigcode/starcoder2-7b}{bigcode/starcoder2-7b} \cite{lozhkov2024starcoder}& 7 & 46.3 \\ \hline
\href{https://huggingface.co/WizardLMTeam/WizardCoder-15B-V1.0}{WizardLMTeam/WizardCoder-15B} \cite{luo2023wizardcoder}& 15 & 59.8\\ \hline
\href{https://huggingface.co/WizardLMTeam/WizardCoder-33B-V1.1}{WizardLMTeam/WizardCoder-33B} \cite{luo2023wizardcoder}& 33 & 79.9 \\ \hline

\end{tabular}
\caption{LLMs considered in this study.}
\label{tab:model_data}
\end{table}

%46.7B (12.9B)
%\footnote{Mixtral has 46.7B total parameters but only uses 12.9B parameters per token. It, therefore, processes input and generates output at the same speed and for the same cost as a 12.9B model \url{https://mistral.ai/news/mixtral-of-experts/}} 
%%%https://evalplus.github.io/leaderboard.html

%% file: figshapes.tex
\begin{figure*}
\centering

            \begin{subfigure}{0.32\textwidth}
                        \centering
                        \includegraphics[width=\textwidth]{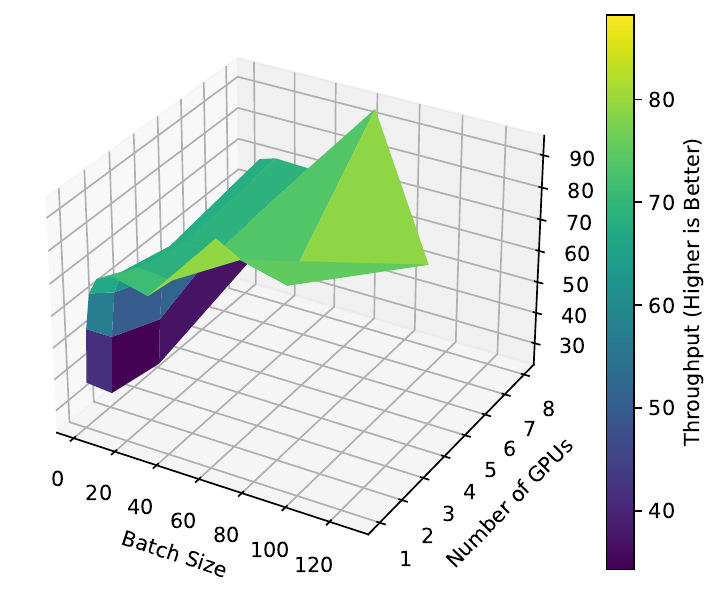}
                        \caption{Bloom-1b7 \hfinference{} \nvidia{} A100}
                \label{fig:plot:space:a100_autohf_Bloom-1b7}
                    \end{subfigure}
             \begin{subfigure}{0.32\textwidth}
                        \centering
                        \includegraphics[width=\textwidth]{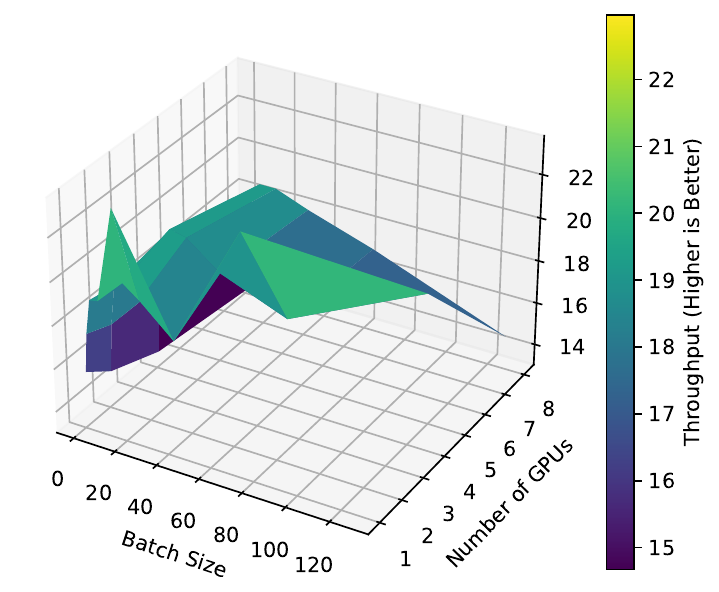}
                        \caption{Codellama-7b \hfinference{} \nvidia{} A100}
                        \label{fig:plot:space:a100_autohf_Codellama-7b-hf}
                    \end{subfigure}
      \begin{subfigure}{0.31\textwidth}
                        \centering
                        \includegraphics[width=\textwidth]{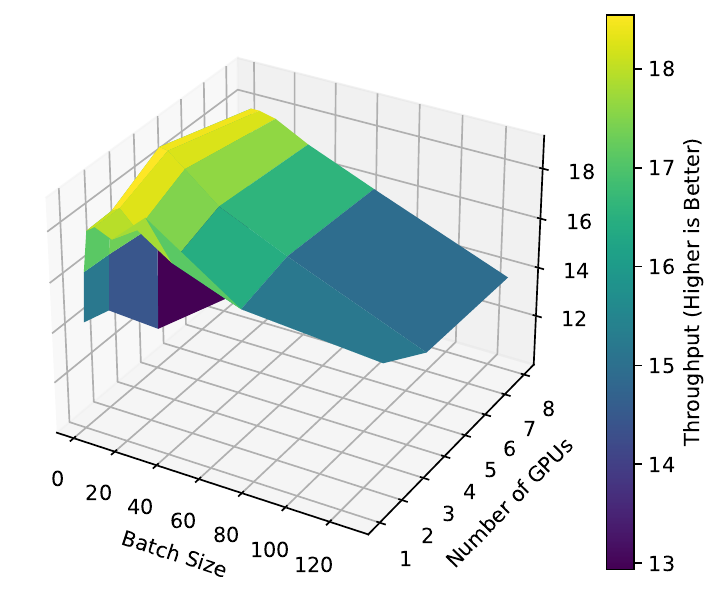}
                        \caption{Mistral-7b \hfinference{} \nvidia{} A100}
                        \label{fig:plot:space:a100_autohf_Mistral-7b}
                    \end{subfigure}

         %  \begin{subfigure}{0.32\textwidth}
        %                \centering
         %   \includegraphics[width=0.9\textwidth]{plots/hyperplane/plot_hyperplane_a100_autohf_Codestral-22b-v0.1_mtg_5_throughput.pdf}
         %               \caption{vLLM-A100}
         %               \label{fig:plot:space:autohf_Codestral-22b}
         %           \end{subfigure}
           \begin{subfigure}{0.32\textwidth}
                        \centering
                        \includegraphics[width=\textwidth]{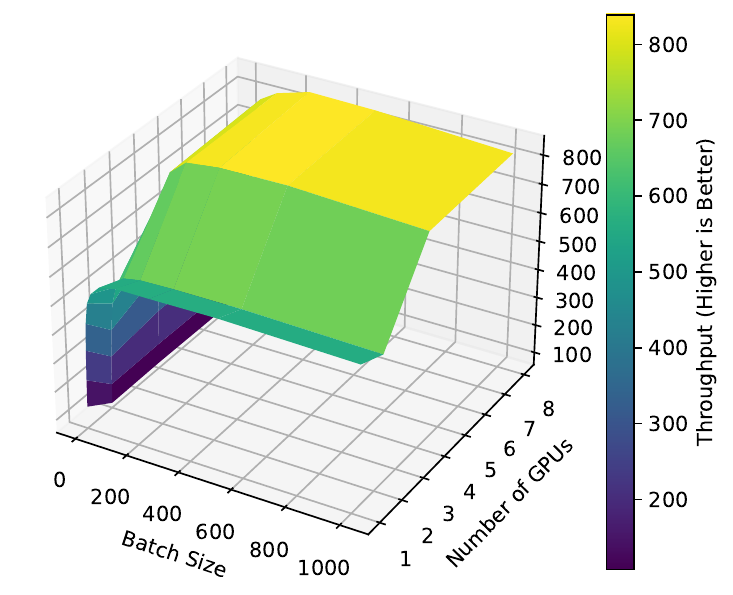}
                        \caption{Gemma-2b \vLLM{} \nvidia{} A100}
                    \label{fig:plot:space:vLLM_Gemma-2b}
                    \end{subfigure}
          \begin{subfigure}{0.32\textwidth}
                        \centering
                        \includegraphics[width=\textwidth]{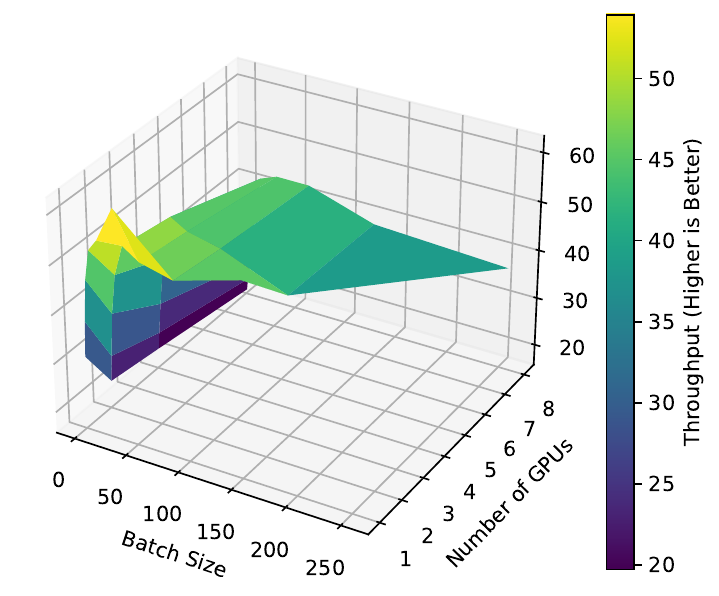}
                        \caption{Starcoder-15b \hfinference{} \nvidia{} A100}
                        \label{fig:plot:space:a100_autohf_Starcoder}
                    \end{subfigure}
  \begin{subfigure}{0.32\textwidth}
                        \centering
                         \includegraphics[width=\textwidth]{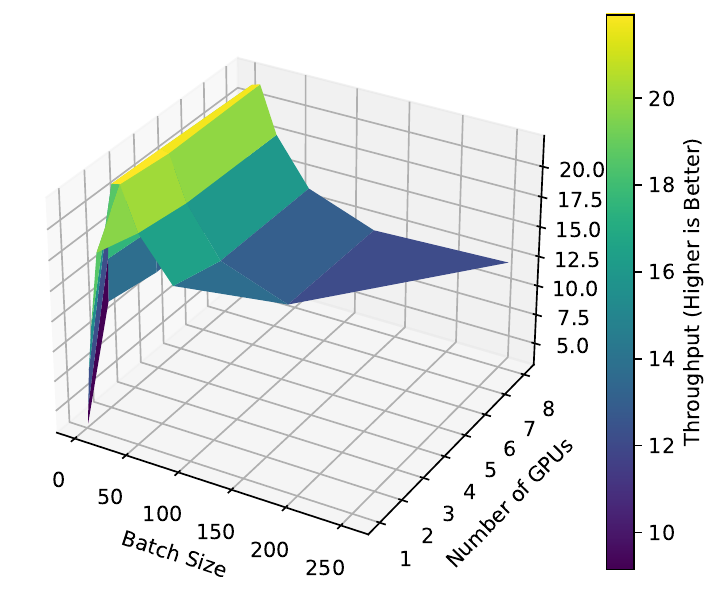}
\caption{Starcoder-15b \hfinference{} \nvidia{} V100}
\label{fig:plot:space:tesla_autohf_Starcoder}
                    \end{subfigure}
                   
\caption{Throughput landscape across the hyperparameter space (batch size and GPUs).}
\label{fig:landscapes}
\end{figure*}
%\begin{figure}
%\includegraphics[width=0.4\textwidth]{plots/hyperplane/plot_hyperplane_tesla_autohf_Starcoder_mtg_5_throughput.pdf}
%\caption{Throughput landscape of StarCoder-15b using a \nvidia{} V100.}
%\label{fig:plot:space:tesla_autohf_Starcoder}
%\end{figure}           

%% file: tables/tabHFvLLMOnline.tex
\begin{table}[t]
\centering
\begin{tabular}{|l||lr|rr||lrr|}
\toprule

& \multicolumn{2}{c|}{\hfinference{}} & \multicolumn{5}{c|}{\vLLM{}} \\ 
\cline{4-8}
& \multicolumn{2}{c|}{(Best)} & \multicolumn{2}{c||}{(Same \#GPUs)} & \multicolumn{3}{c|}{(Best)} \\ 

\llm{} & \rotatebox[origin=c]{90}{GPUs} & Thp & Thp & Impr. & \rotatebox[origin=c]{90}{GPUs} & Thp & Impr. \\
\midrule
Bloom-1b7 & 1 & 37.6 & 133.1 & 3.5X & 1 & 133.1 & 3.5X \\
Bloom-3b & 1 & 27.3 & 98.6 & 3.6X & 1 & 98.6 & 3.6X \\
Bloom-7b1 & 1 & 16.8 & 64 & 3.8X & 2 & 67.2 & 4.0X \\
Codellama-7b & 1 & 15.7 & 63 & 4.0X & 1 & 63 & 4.0X \\
Codellama-13b & 2 & 9.7 & 40.6 & 4.2X & 4 & 48.2 & 5.0X \\
Codellama-34b & 4 & 4.1 & 32.4 & 7.9X & 8 & 34.7 & 8.5X \\
Codellama-70b & 8 & 1.8 & 21.7 & 12.1X & 8 & 21.7 & 12.1X \\
Codestral-22b & 4 & 5.4 & 53.2 & 9.9X & 8 & 64.7 & 12.0X \\
Gemma-2b & 1 & 28.8 & 132.3 & 4.6X & 1 & 132.3 & 4.6X \\
Gemma-7b & 4 & 6.4 & 70.8 & 11.1X & 4 & 70.8 & 11.1X \\
Mistral-7b & 1 & 14.3 & 62.3 & 4.4X & 4 & 66.5 & 4.7X \\
Mixtral-8x7b & 8 & 4.1 & 58 & 14.1X & 4 & 59.2 & 14.4X \\
Phi-1 & 2 & 29 & 73.7 & 2.5X & 1 & 158.5 & 5.5X \\
Phi-1\_5 & 1 & 37 & 171.5 & 4.6X & 1 & 171.5 & 4.6X \\
Phi-2 & 2 & 22 & 68.9 & 3.1X & 1 & 110.3 & 5.0X \\
Starcoder-15b & 1 & 30.7 & 32.5 & 1.1X & 8 & 58.4 & 1.9X \\
Starcoder2-3b & 1 & 44.8 & 104.1 & 2.3X & 1 & 104.1 & 2.3X \\
Starcoder2-7b & 1 & 15.7 & 61.1 & 3.9X & 4 & 62.3 & 4.0X \\
Wizardcoder-15b & 2 & 8 & 45.7 & 5.7X & 8 & 57.3 & 7.2X \\
Wizardcoder-33b & 4 & 1.6 & 36.8 & 23.0X & 8 & 43.8 & 27.4X \\
\bottomrule
\end{tabular}

\caption{Throughput (Thp) from \hfinference{} and \vLLM{} on A100. %\todo{add median/mean}
}
\label{tab:minMaxMetricAutoHFVLLMOnlineInferece}
\end{table}

%% file: figOnlineHFGPUs.tex
\begin{figure}[t]
\centering
                      \begin{subfigure}{0.265\textwidth}
                        \centering
                        \includegraphics[width=\textwidth]{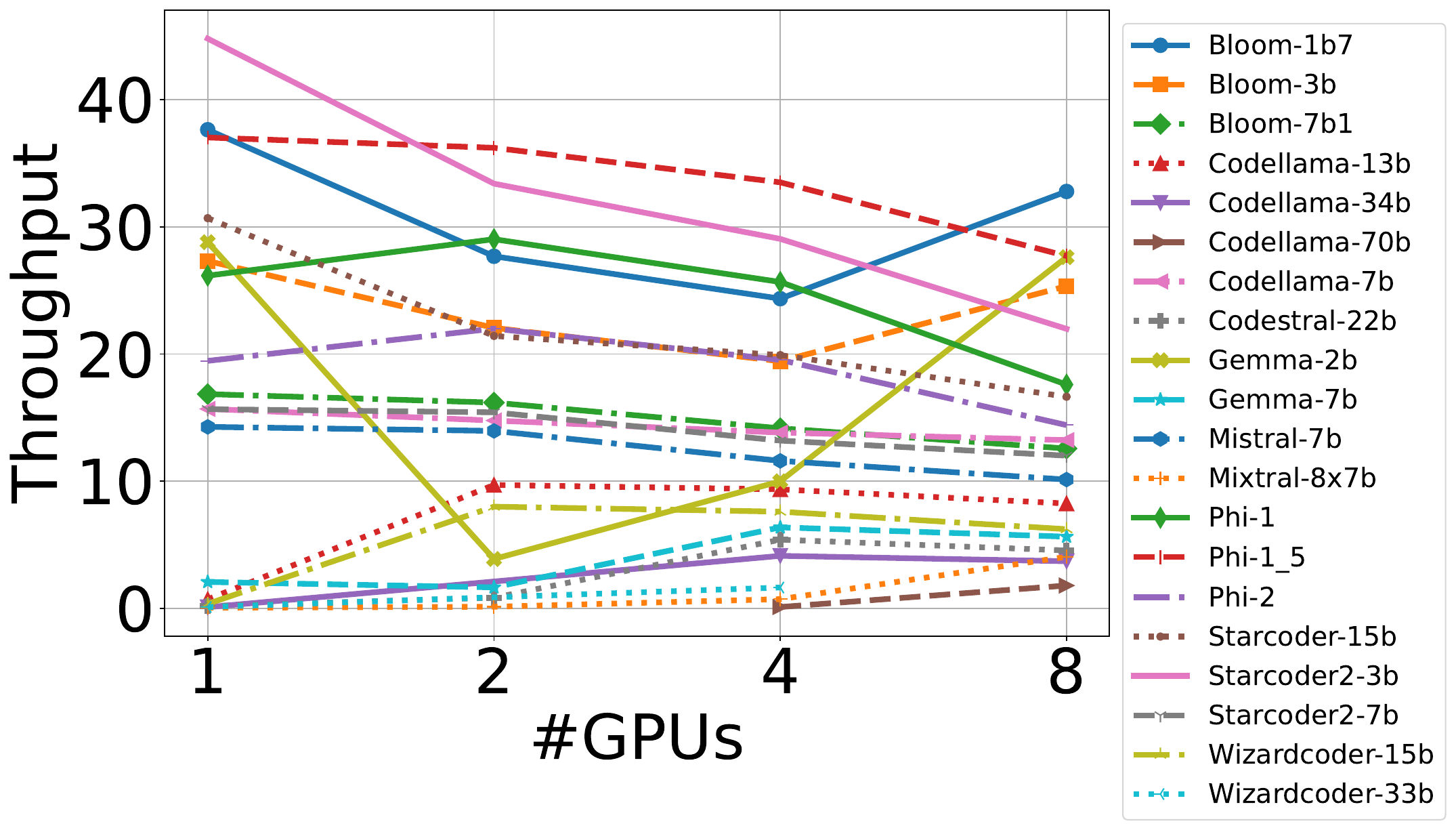}
                        \caption{\hfinference{}}
                        \label{fig:gpus_evol_a100_autohf_throughput}
                    \end{subfigure}
                   \begin{subfigure}{0.215\textwidth}
                        \centering
                        \includegraphics[width=\textwidth]{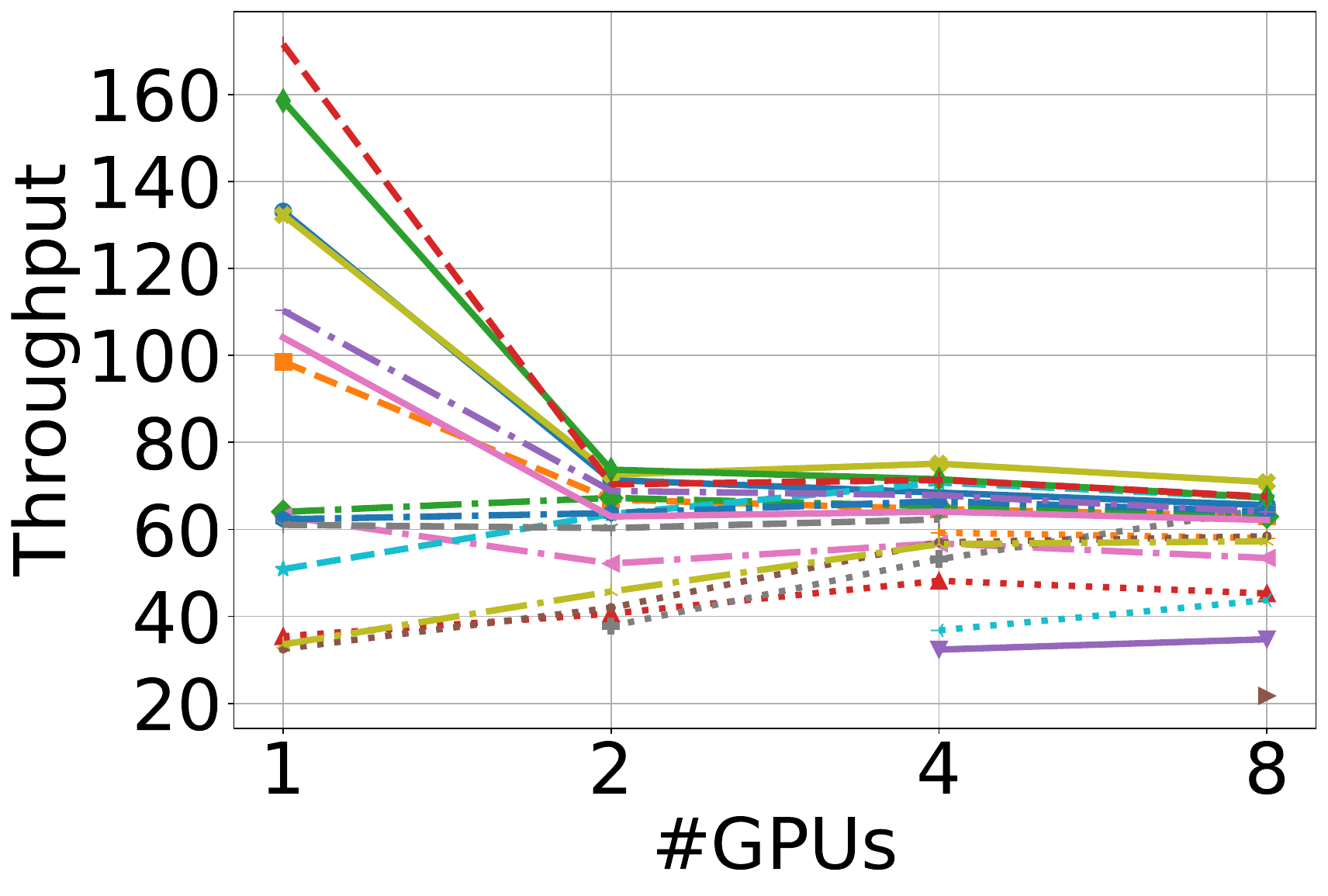}  
                        \caption{\vLLM{}}
                        \label{fig:gpus_evol_a100_vLLM_throughput}
                    \end{subfigure}
                           
\caption{Throughput Variation with Different Numbers of GPUs (\nvidia{} A100) during Online Inference (batch size $=$ 1).}
\label{fig:onlineGPUsAutoHFvLLM}
\end{figure}

%% file: figBatchesAll.tex
\begin{figure*}
\centering

                      \begin{subfigure}{0.245\textwidth}
                        \centering
                        \includegraphics[width=\textwidth]{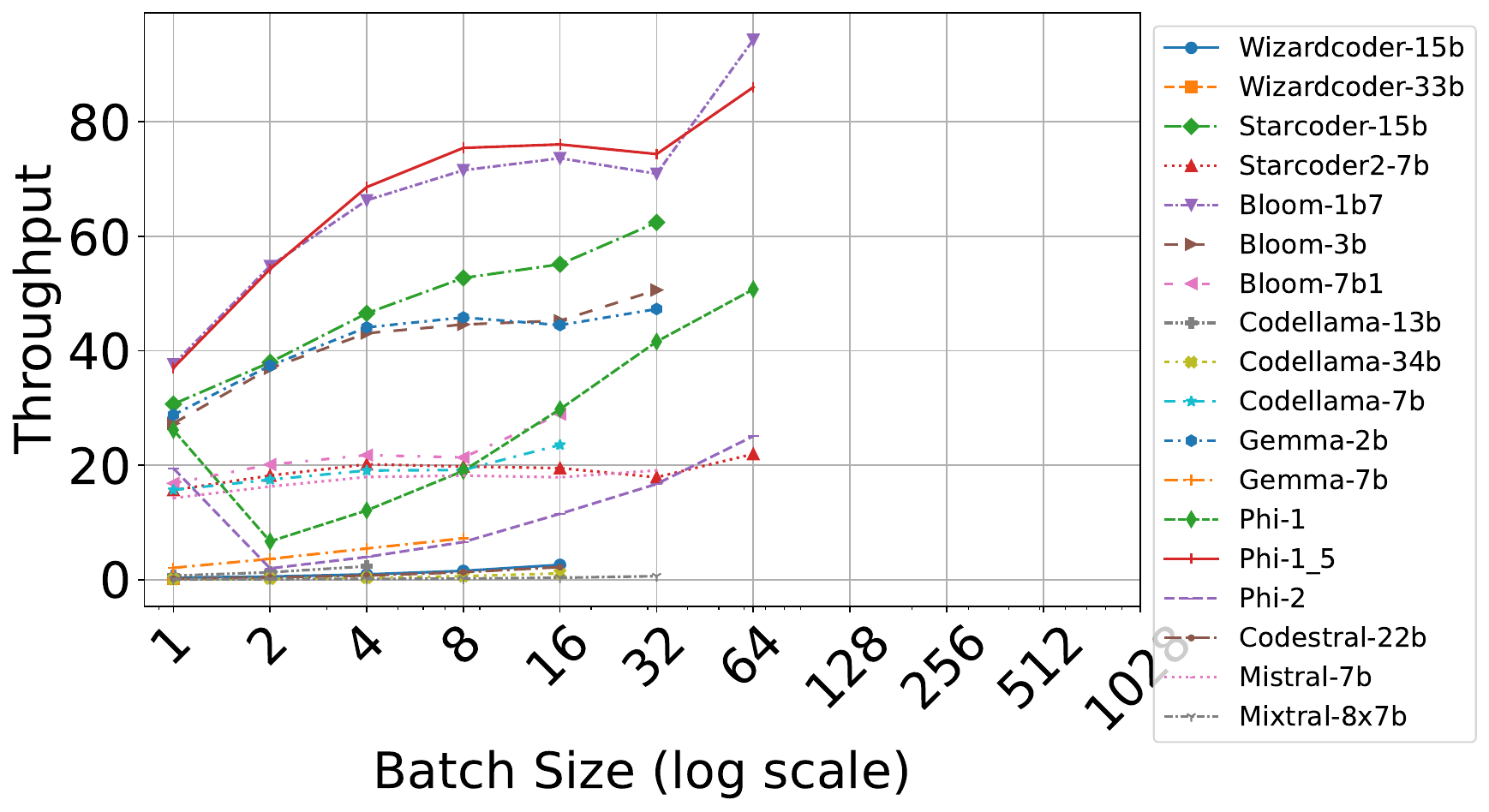}
                        \caption{1 GPU }
                        \label{fig:RQ3_bs_throughput_a100_autohf_1gpu}
                    \end{subfigure}
                    \begin{subfigure}{0.245\textwidth}
                        \centering
                        \includegraphics[width=\textwidth]{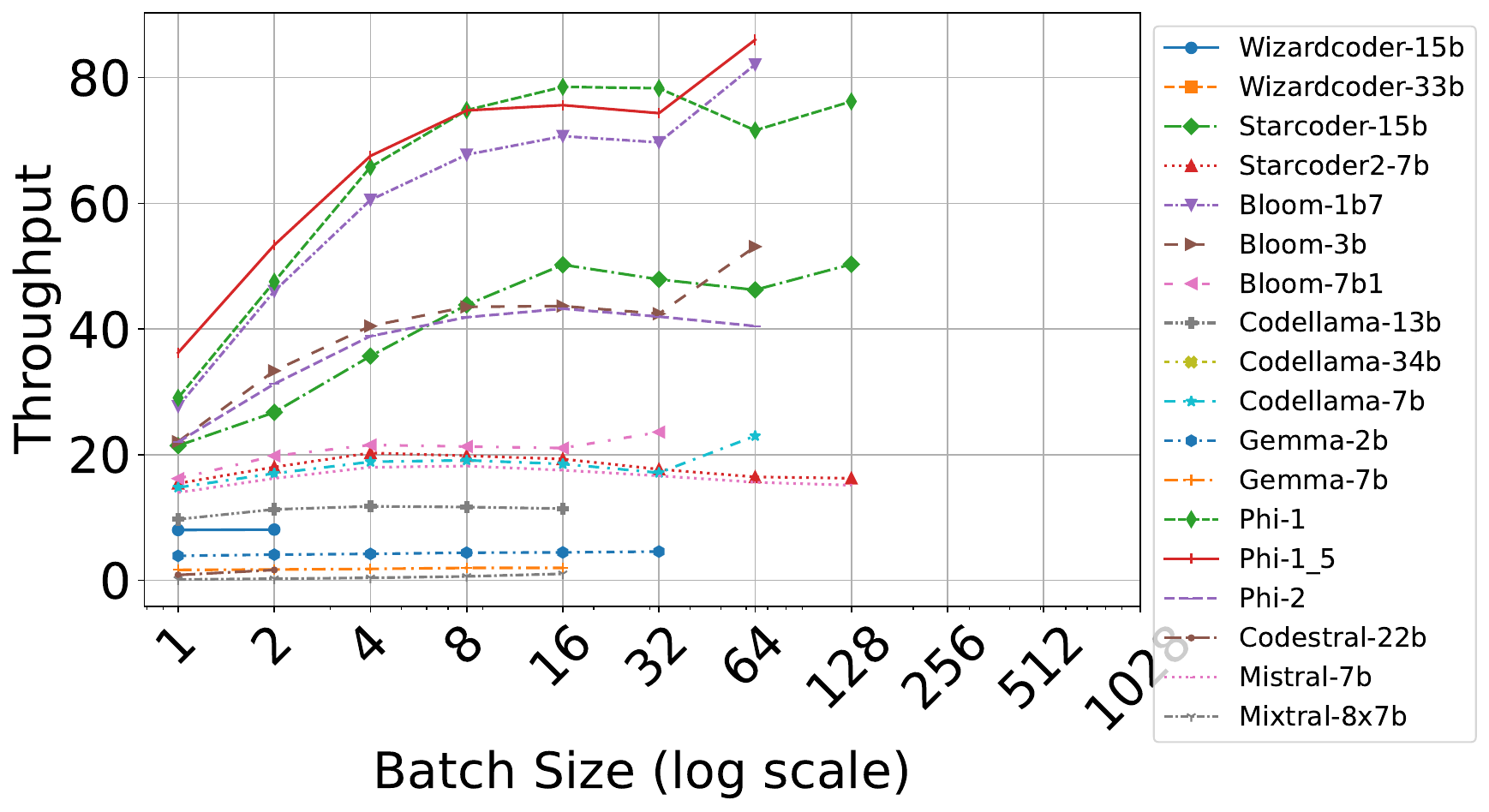}
                        \caption{2 GPUs}
                        \label{fig:RQ3_bs_throughput_a100_autohf_2gpu}
                    \end{subfigure}
                \begin{subfigure}{0.245\textwidth}
                        \centering
                        \includegraphics[width=\textwidth]{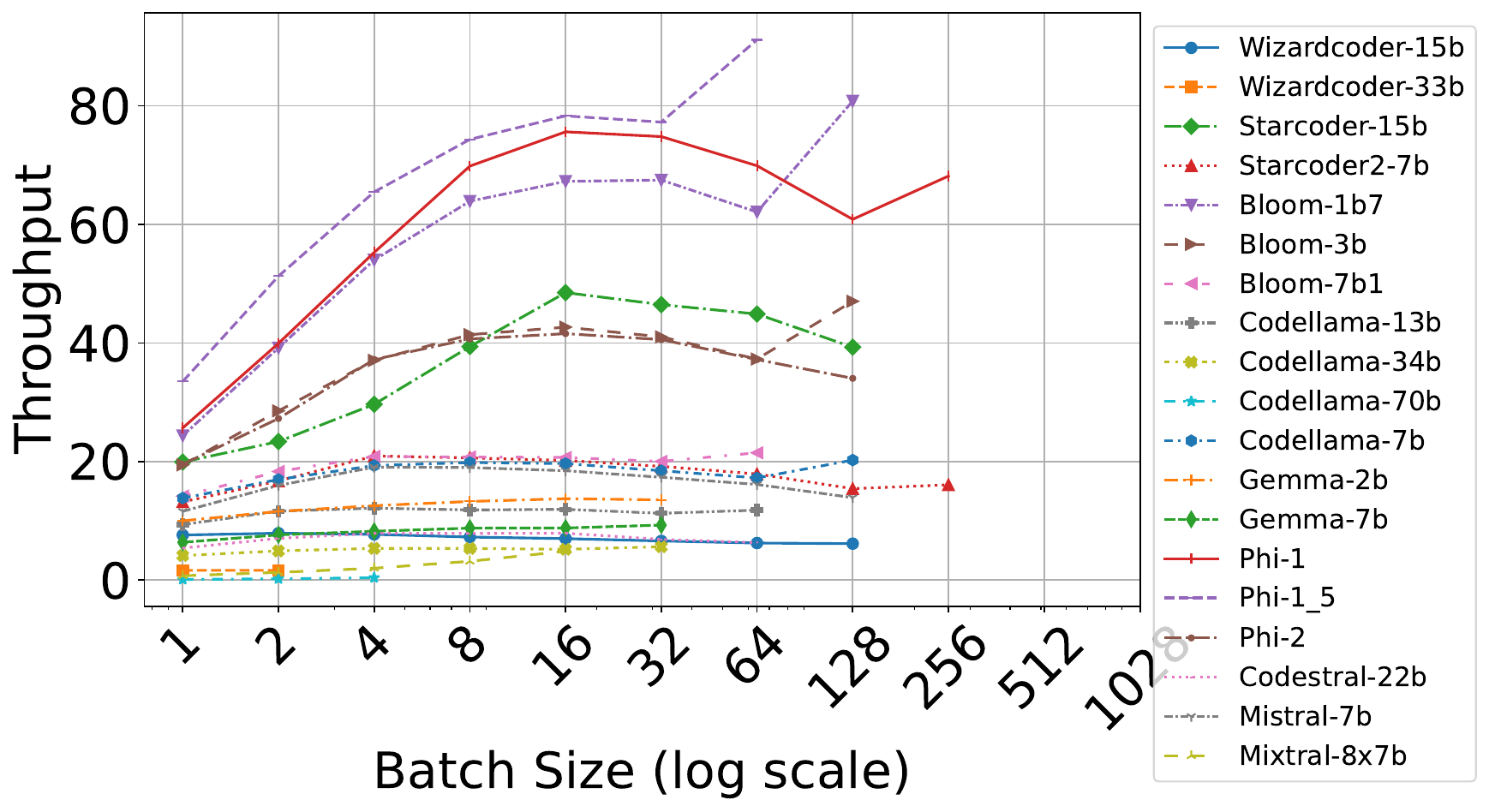}
                        \caption{4 GPUs}
                        \label{fig:RQ3_bs_throughput_a100_autohf_4gpu}
                    \end{subfigure}
                 \begin{subfigure}{0.245\textwidth}
                        \centering
                        \includegraphics[width=\textwidth]{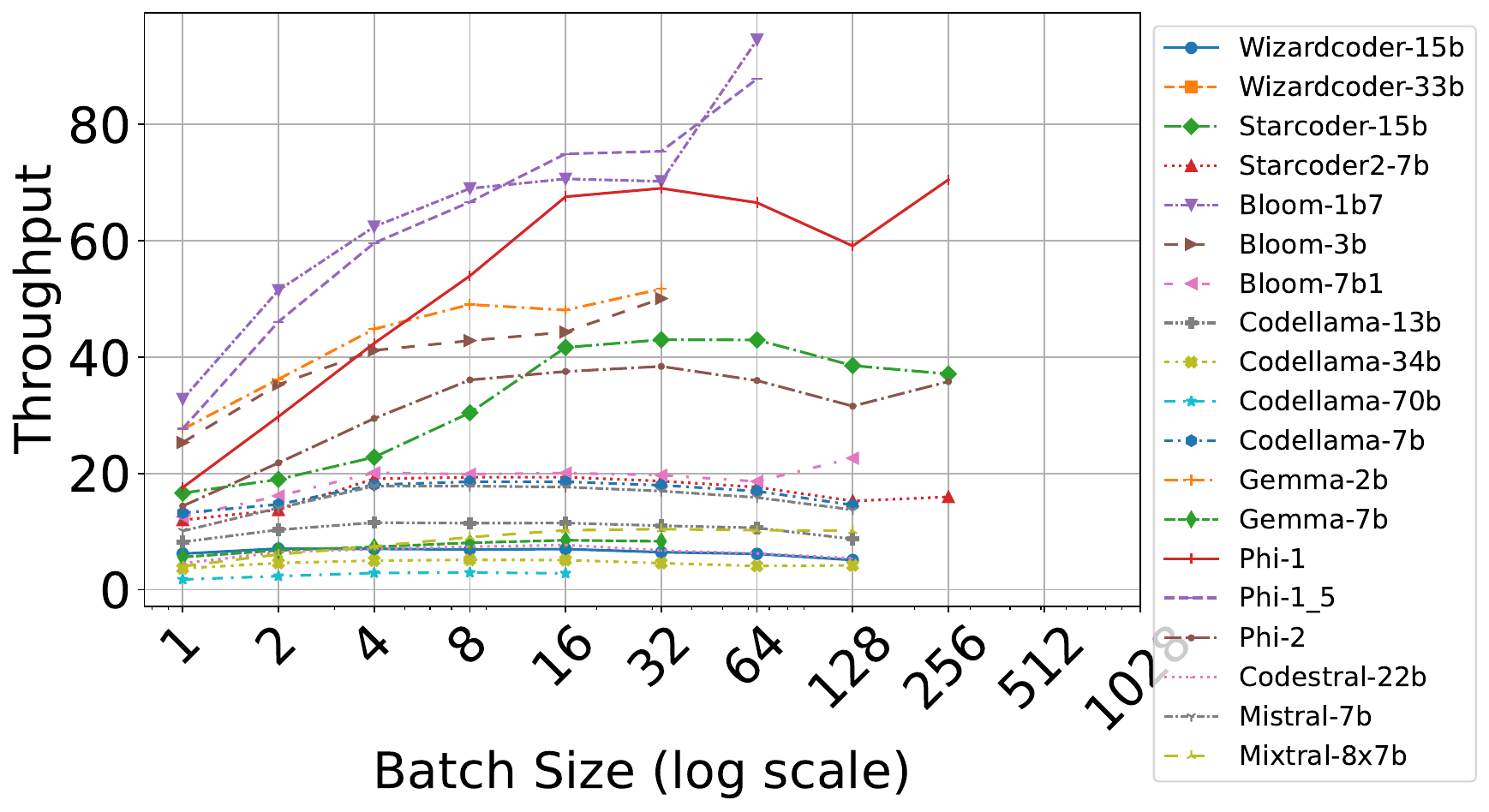}
                        \caption{8 GPUs}
                        \label{fig:RQ3_bs_throughput_a100_autohf_8gpu}
                    \end{subfigure}

\caption{Throughput at difference batch sizes for \hfinference{} using \nvidia{} A1000 GPUs
(outlier Starcoder2-3b removed and shown in \autoref{fig:RQ3_bs_throughput_a100_autohf_all})}
\label{fig:batchesA100}
\end{figure*}

%% file: figBatchSize.tex
\begin{figure}
\centering

                      \begin{subfigure}{0.23\textwidth}
                        \centering
                        \includegraphics[width=\textwidth]{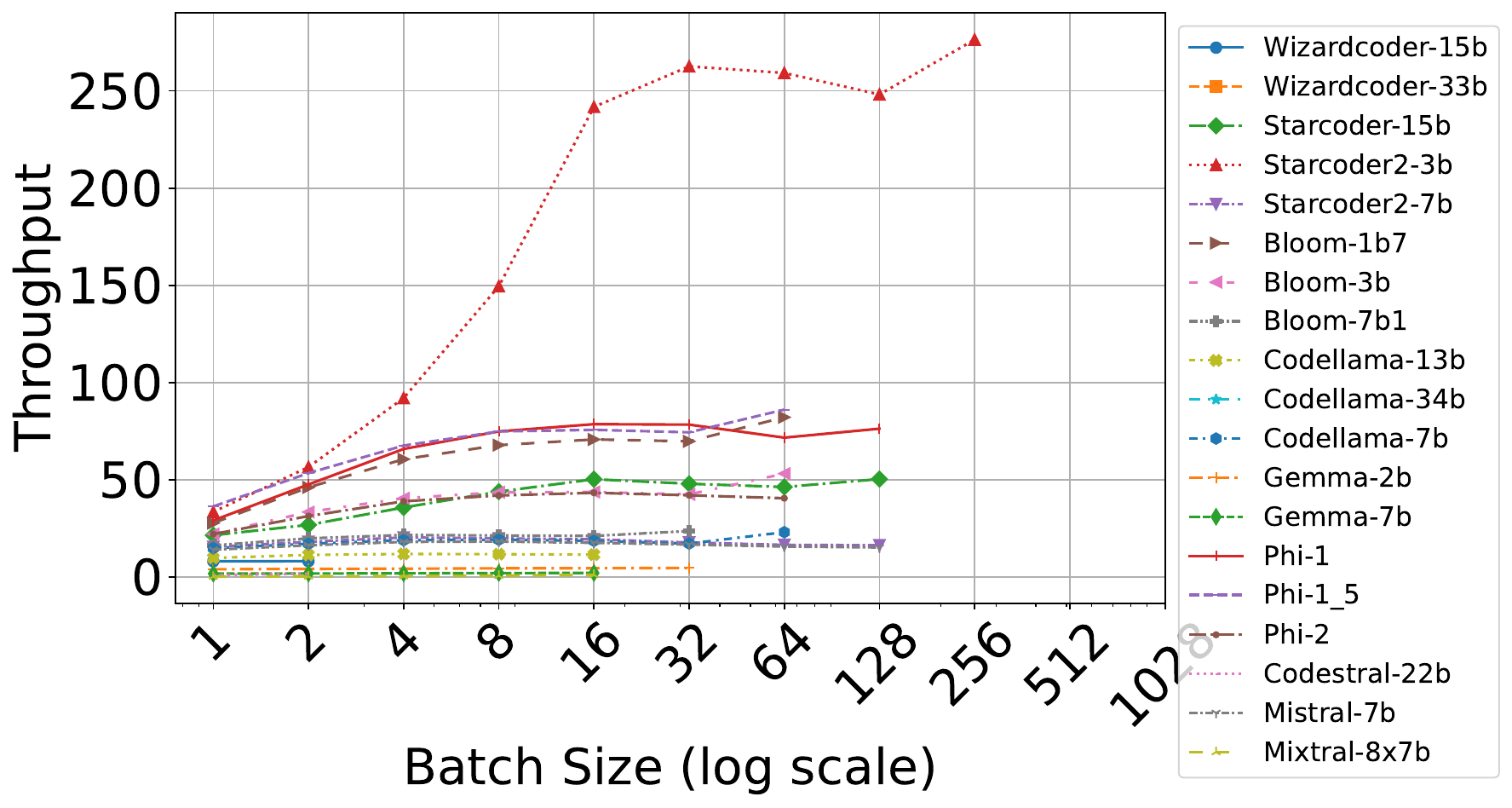}
                        \caption{\hfinference{} }
                        \label{fig:RQ3_bs_throughput_a100_autohf_all}
                    \end{subfigure}
                    \begin{subfigure}{0.23\textwidth}
                        \centering
                        \includegraphics[width=\textwidth]{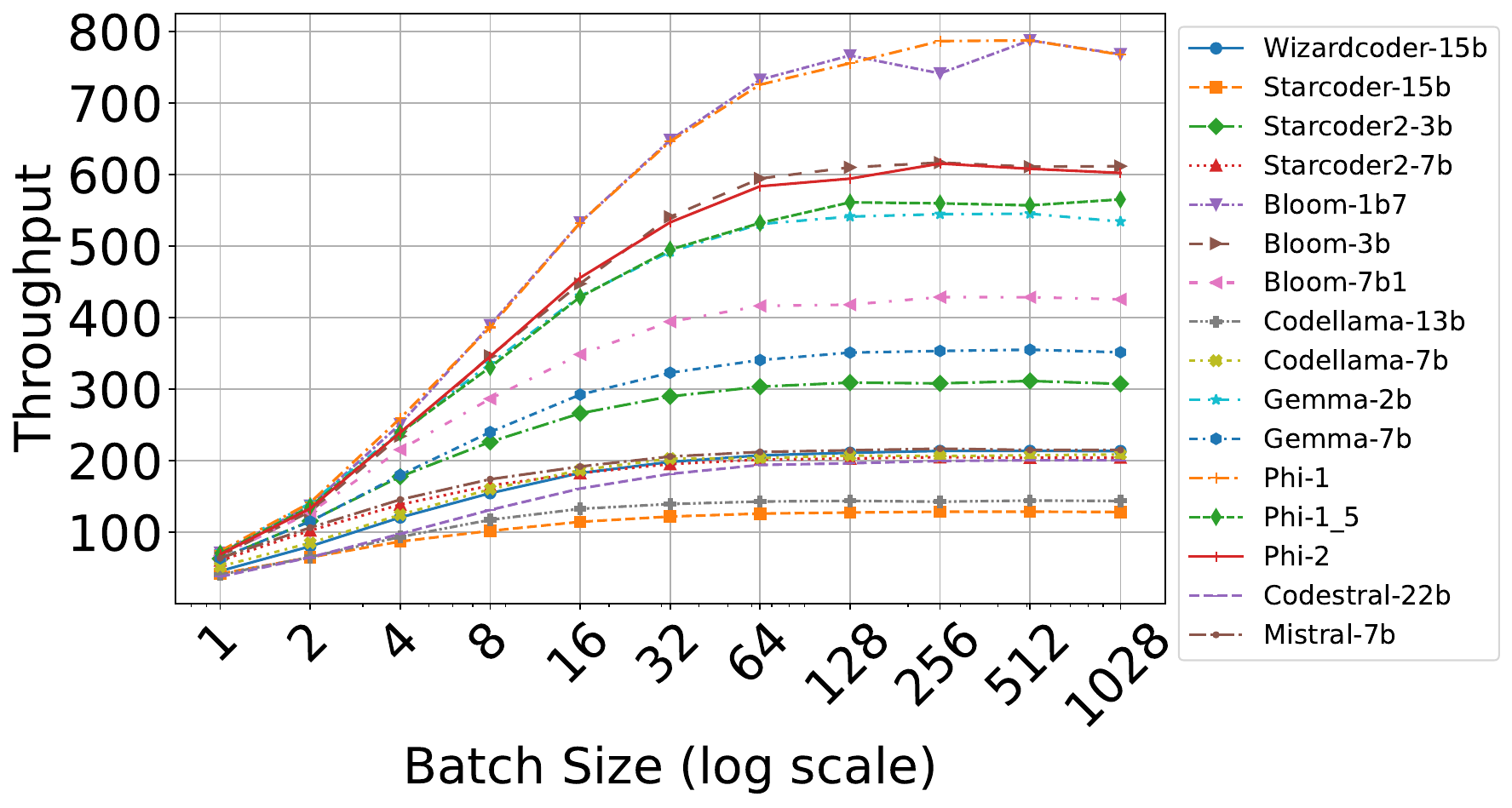}
                        \caption{\vLLM{}}
                        \label{fig:RQ3_bs_throughput_a100_autohf}
                    \end{subfigure}

\caption{Comparison of throughput using two \nvidia{} A1000 GPUs between \hfinference{} (left) and \vLLM{} (right)
}
\label{fig:batchsizes2GPU}
\end{figure}

%% file: figHPO.tex
\begin{figure}
\centering
\includegraphics[width=0.33\textwidth]{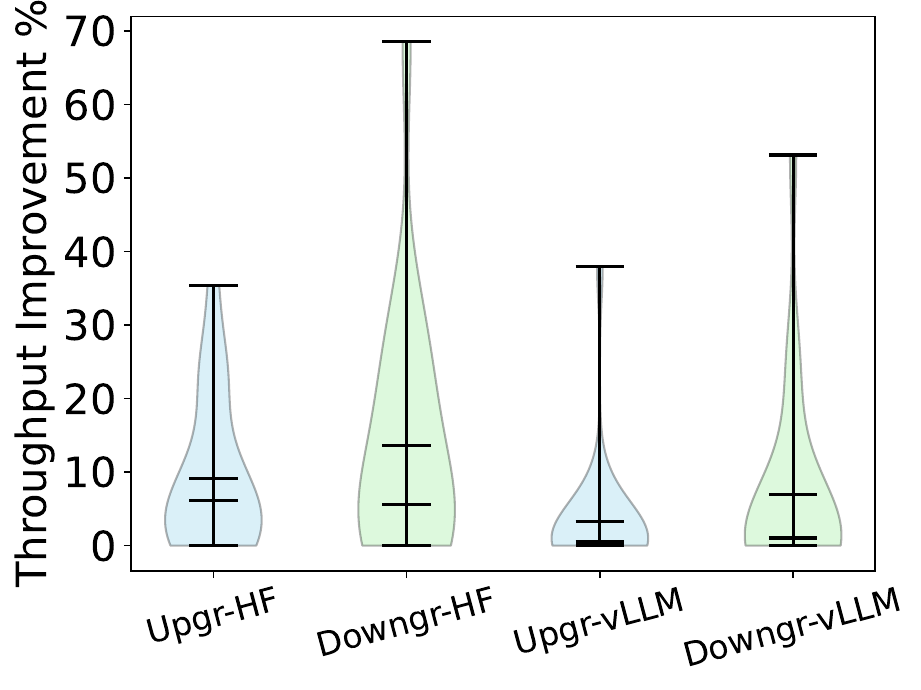}                 
\caption{Distribution of throughput improvement for \hfinference{} given by hyperparameter optimization in hardware upgrading (from \nvidia{} V100 to A100) and downgrading (from A100 to V100).}
\label{fig:violinhpo}
\end{figure}